%% file: main.tex
\documentclass[sigconf]{acmart}

\renewcommand\footnotetextcopyrightpermission[1]{} 
\AtBeginDocument{%
  \providecommand\BibTeX{{%
    \normalfont B\kern-0.5em{\scshape i\kern-0.25em b}\kern-0.8em\TeX}}}

\usepackage{amsmath,amsfonts}
\usepackage{algorithmic}
\usepackage{graphicx}
\usepackage{textcomp}
\usepackage{xcolor}
\usepackage{adjustbox}
\usepackage{tabularx}
\usepackage{arydshln}
\usepackage[euler]{textgreek}
\usepackage{subcaption} 
\usepackage[ruled,vlined]{algorithm2e}
\usepackage[utf8]{inputenc}
\usepackage[english,greek]{babel}
\usepackage{flexisym}
\usepackage{balance} 
\usepackage[colorinlistoftodos]{todonotes}
\usepackage[english]{babel}

\usepackage{enumitem}
\setlist{leftmargin=2mm}

\begin{document}
\fancyhead{}
\selectlanguage{english}
\newcommand{\DLBACALPHA}{\textrm{DLBAC}\textsubscript{\textgreek{a}}}

\newcommand{\DLBACALPHAR}{\textrm{DLBAC}\textsubscript{\textgreek{a}$-$R}}
\newcommand{\DLBACALPHAD}{\textrm{DLBAC}\textsubscript{\textgreek{a}$-$D}}
\newcommand{\DLBACALPHAX}{\textrm{DLBAC}\textsubscript{\textgreek{a}$-$X}}

\newcommand{\SMALLA}{u4k$-$r4k$-$auth11k}
\newcommand{\SMALLB}{u5k$-$r5k$-$auth12k}
\newcommand{\MEDIUMA}{u5k$-$r5k$-$auth19k}
\newcommand{\MEDIUMB}{u4k$-$r4k$-$auth21k}
\newcommand{\MEDIUMC}{u4k$-$r7k$-$auth20k}
\newcommand{\MEDIUMD}{u4k$-$r4k$-$auth22k}
\newcommand{\LARGEA}{u4k$-$r6k$-$auth28k}
\newcommand{\LARGEB}{u6k$-$r6k$-$auth32k}

\newcommand{\AMAZONKAGGLE}{amazon$-$kaggle}
\newcommand{\AMAZONUCI}{amazon$-$uci}

\title{Toward Deep Learning Based Access Control}


\author{Mohammad Nur Nobi}
\affiliation{%
  \institution{Institute for Cyber Security (ICS) and Department of Computer Science}
  \institution{University of Texas at San Antonio}
  \state{Texas}
  \country{USA}
  }
\email{mohammadnur.nobi@utsa.edu}

\author{Ram Krishnan}
\affiliation{%
  \institution{ICS, NSF Center for Security and Privacy Enhanced
Cloud Computing (C-SPECC), and Department of Electrical and Computer Engineering}
  \institution{University of Texas at San Antonio}
  \state{Texas}
  \country{USA}
  }
\email{ram.krishnan@utsa.edu}

\author{Yufei Huang}
\affiliation{%
  \institution{Department of Medicine, University of Pittsburgh, and UPMC Hillman Cancer Center}
  \state{Pennsylvania}
  \country{USA}
  }
\email{yuh119@pitt.edu}

\author{Mehrnoosh Shakarami}
\affiliation{%
  \institution{ICS, C-SPECC, and Department of Computer Science}
  \institution{University of Texas at San Antonio}
  \state{Texas}
  \country{USA}
  }
\email{mehrnoosh.shakarami@my.utsa.edu}

\author{Ravi Sandhu}
\affiliation{%
  \institution{ICS, C-SPECC, and Department of Computer Science}
  \institution{University of Texas at San Antonio}
  \state{Texas}
  \country{USA}
  }
\email{ravi.sandhu@utsa.edu}


\input{contents/abstract}
\maketitle

\input{contents/introduction}

\input{contents/dlbac}
\input{contents/relatedwork}
\input{contents/dlbac-alpha}

\input{contents/evaluation}

\input{contents/interpretability}

\input{contents/discussions}
\input{contents/conclusion}

\begin{acks}
We would like to thank the CREST Center For Security And Privacy Enhanced
Cloud Computing (C-SPECC) through the National Science Foundation (NSF)
(Grant Award \#1736209) and the NSF Division of Computer and Network Systems (CNS) (Grant Award \#1553696) for their support and contributions to this research.
\end{acks}


\balance
\bibliographystyle{ACM-Reference-Format}
\bibliography{main}

\end{document}

%% file: contents/abstract.tex
\begin{abstract}

  A common trait of current access control approaches is the challenging need to engineer abstract and intuitive access control models. This entails designing access control information in the form of roles (RBAC), attributes (ABAC), or relationships (ReBAC) as the case may be, and subsequently, designing access control rules. This framework has its benefits but has significant limitations in the context of modern systems that are dynamic, complex, and large-scale, due to which it is difficult to maintain an accurate access control state in the system for a human administrator. This paper proposes Deep Learning Based Access Control (DLBAC) by leveraging significant advances in deep learning technology as a potential solution to this problem. We envision that DLBAC could complement and, in the long-term, has the potential to even replace, classical access control models with a neural network that reduces the burden of access control model engineering and updates. Without loss of generality, we conduct a thorough investigation of a candidate DLBAC model, called \(\DLBACALPHA\), using both real-world and synthetic datasets. We demonstrate the feasibility of the proposed approach by addressing issues related to accuracy, generalization, and explainability. We also discuss challenges and future research directions.
  
\end{abstract}

\ccsdesc[500]{Security and privacy~Access control}
\ccsdesc[500]{Computing methodologies~Machine learning}

\keywords{Access control; Deep learning; Automation}
  



%% file: contents/introduction.tex
\section{Introduction and Motivation} \label{sec:introduction}

Access Control Lists (ACLs)~\cite{harrison1976HRU}, Role-Based Access Control (RBAC) \cite{sandhu1996rbac}, and Attribute Based Access Control (ABAC)~\cite{hu2013NISTguideToABAC} are some of the mainstream approaches to determine users' access to resources. Commercial solutions \cite{ferraiolo1995commercialACneeds} that cater to organizations employ one or more of these classical access control functionalities.
While tremendous progress has been made in the realm of classical access control approaches~\cite{karp2010ACEvolution},
one fundamental issue has remained the same for over forty years. Skilled security administrators needed to engineer and manage accesses as only humans could develop detailed policy insights about individuals' needs within the broader organization. Clearly, this leads to all types of errors and inefficiencies~\cite{bauer2009real}: there remain plenty of users with accesses that should not have those accesses (over-provisioned to ease administrative burden) and plenty of users that lack accesses that should indeed have those accesses (under-provisioned for the sake of tightened security)~\cite{frank2008, sinclair2008preventative}. Administrators tactfully perform a balancing act to maximize security and minimize costs. This complexity is further exacerbated with the proliferation of cloud-based applications that perform machine-to-machine access through APIs, IoT, BYOD, etc.


In this paper, we propose an automated and dynamic access control mechanism leveraging advances in deep learning technology~\cite{schmidhuber2015deep} that could complement or potentially replace the human administrator. This approach, denoted as Deep Learning Based Access Control (DLBAC), addresses three major limitations of classical access control approaches such as RBAC and ABAC. 
%
Without loss of generality, we use the term attribute to refer to any form of traditional access control information such as roles and relationships.

\textbf{1. Attribute Engineering.}  An organization typically holds a vast number of metadata about its users and resources. However, those metadata are often not meaningful \emph{access control attributes}. As a first step, using organizational context often inferred from those metadata, administrators engineer access control specific attributes that could be used to express access control rules subsequently. This is at best an art today involving semi-formal design and requirements engineering processes~\cite{sainan2019semi-formal}. 

\textbf{2. Policy Engineering.} After access control relevant attributes are engineered, administrators need to engineer access control rules. This is accomplished through either a manual engineering process akin to attribute engineering above or automated mining techniques that take as input a primitive form of access rules such as ACLs and generate approximate ABAC rules (or user-role assignments, in the case of RBAC)~\cite{krautsevich2013policyEngineeringusingRisk, das2018}. We will show that, for complex situations, DLBAC generally captures the access control state of the system with more precision than other approaches which are based on policy mining and classical machine learning.

\textbf{3. Generalization.} Most prior approaches~\cite{narouei2019natureInspiredABACMining, bui2019greedyforRebacMining} that mine access control rules from simpler forms of access control states such as ACLs focus on accurately capturing the access control state as given in those ACLs. Unfortunately, that leads to poor generalization~\cite{cotrini2018Rhapsody, xu2014mining}---that is, the ability to make better access control decisions on users and resources with attributes that were not explicitly seen during the mining process. However, this is something machine learning methods, especially deep learning, are better at. They have an innate ability to make quality predictions as long as the test sample at prediction time aligns with the training data \emph{distribution}. We will show that the engineered rules typically make poor access control decisions for user-resource metadata that were not explicitly seen by the mining process.

DLBAC addresses the above issues by exploring a fundamentally different approach to how access control is designed today.
As illustrated in Figure~\ref{fig:decisionMaking}, DLBAC differs from classical approaches by making decisions based on the \emph{metadata} of users and resources and a trained neural network. (The key distinction between the notions of \emph{metadata} and \emph{attributes} albeit semantic has important practical benefits, which is explained in section~\ref{sec:decisionMakingInClassicalDLBAC}.) It accomplishes this (see Figure~\ref{fig:conceptualRepresentation}) by first replacing access control policies with a neural network that instead makes access control decisions. Second, the neural network is trained using raw metadata from the organization instead of laboriously engineered access control attributes. 

To summarize, we make the following contributions to the field of access control:
\begin{itemize}
    \item We propose DLBAC, a new approach of automated and dynamic next generation access control.
    \item We develop a candidate DLBAC model, \DLBACALPHA, which outperforms classical policy mining and machine learning techniques in many aspects, including capturing the existing access control state of the system accurately and generalizing well to situations that were not seen during training time. 
    \item As DLBAC is a neural network, we address previously highlighted concerns on the explainability of the black-box nature of neural network-based systems for access control~\cite{cappelletti2019quality}. We apply deep learning interpretation methods to confirm that decision-making in DLBAC can indeed be understood to a large degree (albeit not with 100\% accuracy).
    
    \item We synthesize several large-scale access control datasets with a varying number of users and resources. We evaluate the performance of DLBAC on those synthetic datasets along with two real-world datasets.
    
\end{itemize}




The rest of the paper is organized as follows. Section~\ref{sec:dlbac} presents an overview of the DLBAC approach. We discuss related work in Section~\ref{sec:relatedwork}. Section~\ref{sec:dlbac-alpha} introduces some real-world datasets and presents the synthetic data generation method for \(\DLBACALPHA\), a candidate DLBAC model. In the same section, we also explain the implementation of \(\DLBACALPHA\). We conduct performance evaluation of \(\DLBACALPHA\) in Section~\ref{sec:evaluation}. We present approaches to understand \(\DLBACALPHA\) decisions in Section~\ref{sec:understandability}. In Section~\ref{sec:discussion}, we discuss future research directions, and conclude in Section~\ref{sec:conclusion}.



%% file: contents/dlbac.tex
\section{Deep Learning Based Access Control} \label{sec:dlbac}

\begin{figure}[t]
\vspace{-2ex}
\centering
	\includegraphics[width=\linewidth]
	{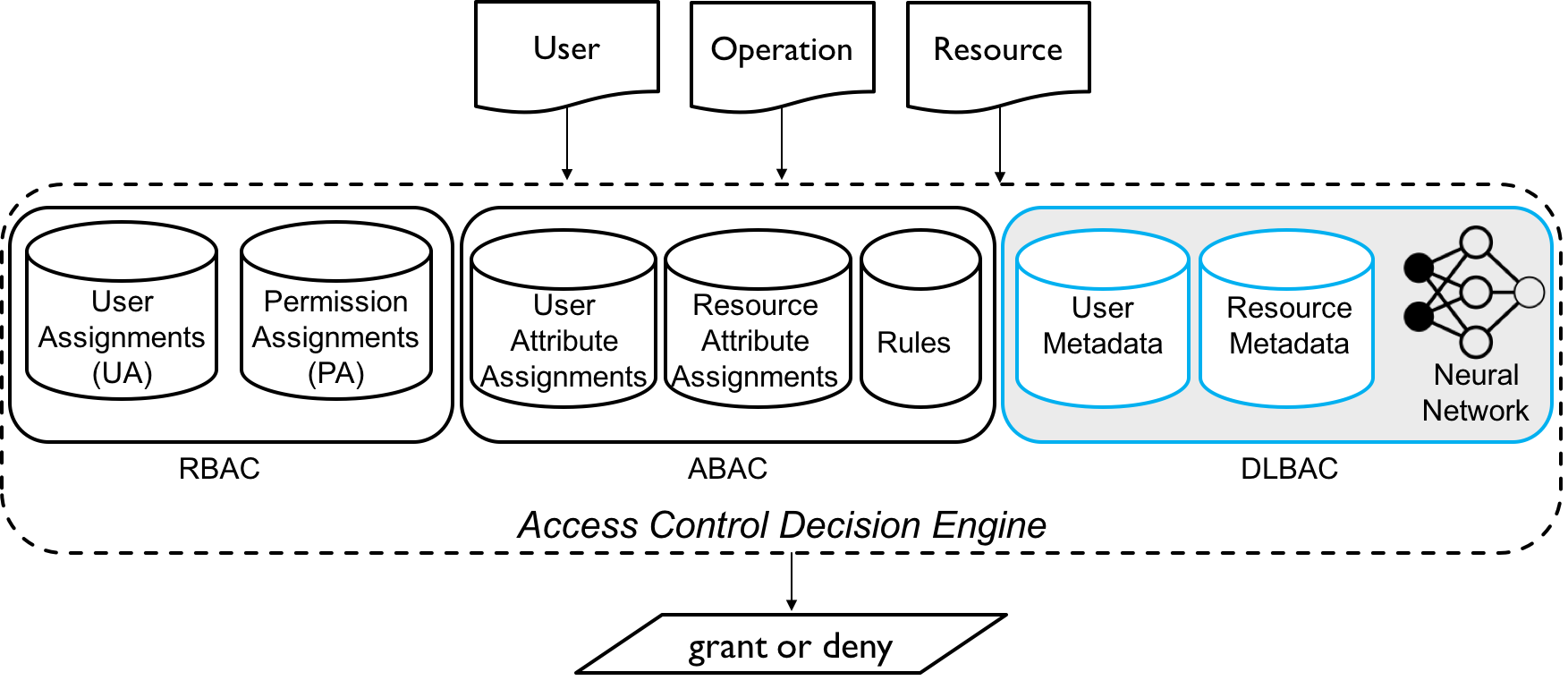}
	\caption{Decision Making in Classical Approaches vs. DLBAC.}
	\label{fig:decisionMaking}
	\vspace{-2ex}
\end{figure}

In this section, we provide a brief overview of DLBAC and explain how it differs from classical approaches.

\subsection{Decision Making in Classical Approaches vs. DLBAC}
\label{sec:decisionMakingInClassicalDLBAC}
Figure~\ref{fig:decisionMaking} illustrates how DLBAC makes a decision as compared to two classical access control approaches (the discussion applies to other forms of access control approaches such as relationship based access control or ReBAC~\cite{cheng2016}). In RBAC, an access control decision is simply a cross-reference between user-role and permission-role assignment relations. In the case of ABAC, an access control rule is evaluated for a given operation based on the attributes of the user and resource in question (sometimes attributes of other entities such as ``environment'' are used as well). In DLBAC, a deep neural network makes an access control decision based on the available \emph{metadata} for the user and resource.
For example, metadata could include logs of accesses, employee join date, access time, 
network access profile
, etc. For simplicity, we assume metadata are represented as name-value pairs. While syntactically they appear to be the same as attributes, which are often name-value pairs as well, semantically they are very different. Metadata are primarily different from attributes since they do not go through the access control design and engineering process. A typical organization could host multiple applications such as email, file storage, human resources, benefits, and other cloud services. Each of those applications hold metadata about users and resources in the organization. Metadata are designed inherently as part of the functionality engineering phase of the system instead of during the access control design phase of the system. Therefore, they are immediately available to DLBAC once the system is implemented. 
For example, `join\_date', `spending\_history' and  `credit\_history' could be metadata of customer, whereas an engineered attribute could be `status' (such as `status $=$ platinum'), determined based on all of those metadata.



\begin{figure}[t]
\vspace{-2ex}
\centering
	\includegraphics[width=\linewidth]
	{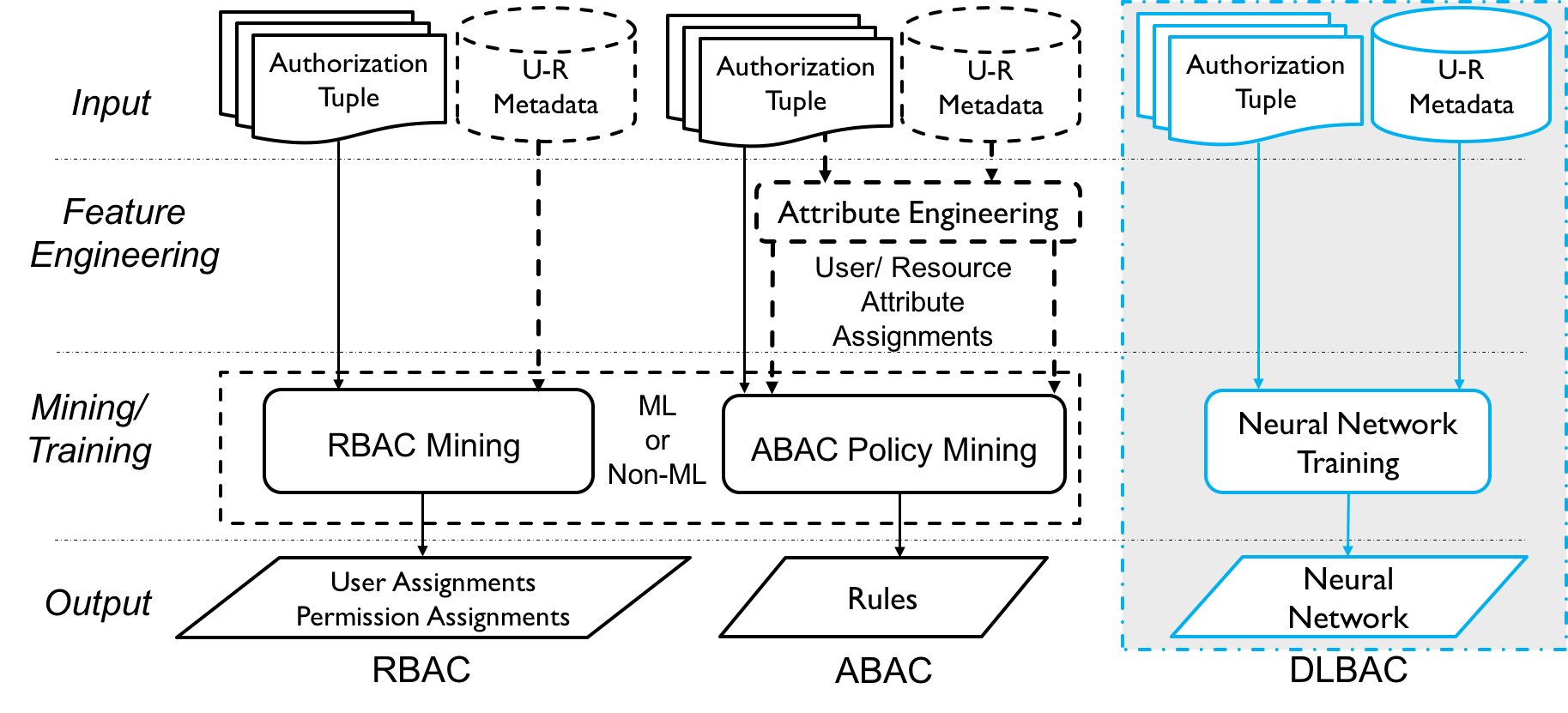}
	\caption{Design Process of Classical Approaches vs. DLBAC.}
	\label{fig:conceptualRepresentation}
	\vspace{-2ex}
\end{figure}

\subsection{Policy Engineering in Classical Approaches vs. DLBAC}
\label{sec:policyEngineeringInClassicalApproaches}

A conceptual representation of classical access control approaches versus DLBAC is depicted in Figure~\ref{fig:conceptualRepresentation}. For simplicity, we assume all methods obtain the current access control state of the system as authorization tuples (e.g., $\langle$user, resource, operations$\rangle$), and the metadata of users (e.g., $\langle$designation, ``employee''$\rangle$) and resources (e.g., $\langle$size, ``small''$\rangle$) as the input.
In ABAC, the first step of \textit{attribute engineering} involves designing users and resources attributes in the system that are selected and properly assigned based on available metadata. Common to ABAC and RBAC, the second stage is \textit{policy mining/engineering}, through which proper policies are developed.  Access control mining algorithms, including those using machine learning (ML), are summarized in Section~\ref{sec:relatedwork}. The last element in the conceptual representation of models is the output. For the RBAC approach, the mining process's output is a set of roles, permission assignment to roles (PA), and user assignment to roles (UA). For ABAC, the output includes a policy consisting of a set of access control rules. (Note that most ABAC mining works assume that attributes and attribute assignments are already available~\cite{sanders2019miningLeastPriviledgeABAC,narouei2019natureInspiredABACMining}). In contrast, DLBAC is an end-to-end access control approach. It does not need attribute engineering since it works directly with the users and resources metadata. The DLBAC approach's output is a trained neural network, which takes user and resource metadata as input and makes access control decisions. We note that the DLBAC is agnostic to any deep neural network architecture. 

%% file: contents/relatedwork.tex
\section{Related Work} \label{sec:relatedwork}
There is plenty of work on mining/engineering policies either by analyzing the current access control state in the forms of logs and ACLs or transforming one access control model to another. We review three classes of related work below.

\textbf{Classical Policy Mining Approaches.}
A rich body of research on mining classical access control models includes RBAC, ABAC, and ReBAC.
Many algorithms were proposed to mine RBAC policies~\cite{xu2012, frank2013}, following either top-down~\cite{vaidya2010, molloy2009, mitra2016, das2018} or bottom-up~\cite{schimpf2000} methods,
while hybrid methods~\cite{frank2009} combine the advantages of both approaches. 
Researchers proposed different criteria to assess the quality of mined policy~\cite{molloy2009} or satisfy various constraints while mining~\cite{jafarian2015}.
Xu and Stoller proposed an ABAC mining algorithm~\cite{xu2014mining}, a variant of which was developed to mine ABAC policies from logs~\cite{xu2014miningFromLogs}. Many approaches have been proposed to mine ABAC policies---for instance, multi-objective evolutionary optimization framework~\cite{medvet2015}, identifying functional dependencies in database tables to mine ABAC policies~\cite{talukdar2017}, and algorithms to find both positive and negative policy rules~\cite{iyer2018mining}. 
Bui et. al. \cite{bui2017} utilized XuStoller's algorithm for mining policies in the ReBAC~\cite{cheng2016} context. Extended versions of this research were presented in~\cite{bui2018,bui2019rebac}. They also proposed a greedy approach for mining ReBAC policy~\cite{bui2019greedyforRebacMining}.
Iyer et al.~\cite{iyer2019} proposed a ReBAC mining algorithm in evolving systems for mining graph transitions. The authors later proposed a method for active learning~\cite{iyer2020active} of ReBAC policies from a black-box access control decision engine using authorization and equivalence queries. A universal access control policy mining, called Unicorn, was proposed by Cotrini et al. ~\cite{cotrini2019Unicorn}, which builds policies in a class of access control models including RBAC, ABAC, and ReBAC.

\textbf{Using Machine Learning (ML) for Mining Access Control.}
Many researchers applied ML algorithms for mining access control policies.  A probabilistic model for the role mining problem driven from the logical structure of RBAC was proposed by Frank et al.~\cite{frank2008}. 
Deep learning was used to identify relevant attributes~\cite{alohaly2018} to mine ABAC policies from natural language. 
Other methods including classification trees~\cite{chari2016}, deep recurrent neural network (RNN)~\cite{narouei2017}, K-Nearest Neighbor (KNN)~\cite{Elhadje2017}, Decision Tree~\cite{xiang2019towards,bui2020learningReBAC-ABAC,bui2020decisionTree} and Restricted Boltzmann Machine (RBM) model~\cite{Mocanu2015} have also been used to mine ABAC policy. The first unsupervised learning-based ABAC mining method used k-modes clustering to mine rules from historical operation data~\cite{karimi2018}.
Naroui et al. ~\cite{narouei2019natureInspiredABACMining} proposed to improve an existing ABAC policy by mining a policy using ML.
Cotrini et al.~\cite{cotrini2018Rhapsody} proposed an ABAC policy mining algorithm, named Rhapsody, built upon APRIORI-SD~\cite{kavvsek2006apriori-sd} which is an ML algorithm for subgroup discovery. Rhapsody mines a generalized policy based on sparse logs. 
Jabal et al.~\cite{jabal2020polisma} proposed a novel framework for learning ABAC policies from data, named Polisma, combining data mining, statistical, and ML techniques. The Polisma mine a set of rules and applies ML techniques to include requests not covered by the mined rules. Karimi et al.~\cite{karimi2021automatic} proposed an automatic approach for learning ABAC policy 
by extracting rules containing both positive and negative attributes and relationship filters. 

\textbf{Classical ML for Access Control Decision Making.}
Classical ML approaches have been widely used for making access decisions.
Sanders et al.~\cite{sanders2019miningLeastPriviledgeABAC} presented an approach to mine ABAC policies while satisfying the least privilege principle in a large-scale organization. Symbolic and non-symbolic ML algorithms to infer ABAC policies from access logs have been proposed by Cappelletti et al.~\cite{cappelletti2019quality}.  
Liu et al. \cite{liu2021efficient} proposed a permission decision engine scheme for ABAC based on Random Forest~\cite{breiman2001random}. The method, called EPDE-ML, decoupled the decision engine from the real access control policy by moving policy updates into a separate phase.

Following our discussion in Section~\ref{sec:dlbac}, DLBAC is fundamentally different from current approaches. It aims to replace traditional rule-based methods with a neural network, leading to better decision accuracy, generalizability, and engineering ease. Comparing to the related works that uses classical machine learning, we will show that a deep learning based approach provides superior performance while being amenable to usable explainability in practice.

%% file: contents/dlbac-alpha.tex
\section{\(\DLBACALPHA\): A Candidate DLBAC Model} \label{sec:dlbac-alpha}

This section presents a prototype of DLBAC, namely \(\DLBACALPHA\), which is an access control model built upon the proposed DLBAC approach in Section~\ref{sec:dlbac}.
As illustrated in Figure~\ref{fig:conceptualRepresentation}, DLBAC models e.g., \(\DLBACALPHA\) need to be fed with authorization tuples and user/resource metadata. 
We apply \(\DLBACALPHA\) to two real-world and eight synthetic datasets. First, we explain synthetic datasets construction and introduce real-world datasets. Then, we discuss how the \(\DLBACALPHA\) neural network is trained, and access decisions are made.

\subsection{Synthetic Dataset Generation}
\label{sec:datasetGeneration}
An approach to generate synthetic access control datasets was proposed in \cite{xu2014mining}. We adopt this approach with minor changes to generate multiple synthetic datasets and briefly discuss here.
The algorithm first generates a set of attribute names for users and resources randomly. Next, it generates a set of \textit{rules} based on those attributes and then uses these rules to inform user/resource creation and attribute value assignments. Each \textit{rule} is a tuple of the form $\langle \mathit{UAE}; \mathit{RAE}; OP; C \rangle$, where $\mathit{UAE}$ is the set of user attribute expression, $\mathit{RAE}$ is the set of resource attribute expression, $C$ is the set of constraints, and $OP$ is a set of operations. For example,
\textit{$\langle$title=student; type=document; {read}; department=department$\rangle$} is a rule where \textit{title=student} represents the UAE, \textit{type=document} is the RAE, \textit{read} is the operation, and \textit{department=department} is the constraint.
A user will be authorized to operate on a resource if the user satisfies the UAE, the resource satisfies the RAE, and both the user and resource satisfy the constraint. 
For each rule, the algorithm generates a set of users that satisfy the rule and then generates resources where for each resource, there is at least one user available to satisfy the rule. 
The following user and resource are created based on the above rule: 
\textit{user(student1, title=student, department=cs)}, 
\textit{resource(document1, department=cs, type=document)}. 
Here, the \textit{student1} and \textit{document1} are the unique ids of a user and a resource, respectively. The user \emph{student1} has two attributes (\emph{title} and \emph{department}), and the resource \emph{document1} has two attributes (department and type). Also, student1 satisfies UAE, as the user has the title \textit{student} which is part of the title in UAE. Similarly, document1 satisfies RAE for the \emph{type} attribute. Both \emph{student1} and \emph{document1} satisfy the constraint as they are from the same department. Thus, according to this rule, \emph{student1} has \emph{read} access to \emph{document1}.

Finally, once the rules are generated, users and resources are created, and attributes are assigned, it is straight-forward to create the authorization tuples. For each user, resource and operation combination that satisfies a rule, an authorization tuple is created or updated with a new operation, as the case may be.

\subsubsection{Syntax of Synthetic Dataset}
We adapt this data generation approach by creating many metadata instead of attributes.
We maintain four operations and various metadata (eight to thirteen) for each user/resource for different datasets.
We define the syntax of \(\DLBACALPHA\)'s dataset to contain a set of authorization tuples. An authorization tuple could be illustrated of the form $\langle uid | rid | m^u_1:v_1, m^u_2:v_2, ..., m^u_i:v_i | m^r_1:v_1, m^r_2:v_2, ..., m^r_j:v_j | \langle op_1, op_2, op_3, op_4\rangle\rangle$. The uid and rid in the tuple indicates the unique id of a user and a resource, respectively. The next part gives the metadata values of all $i$ metadata of a user and $m^u_1$ indicates the first user metadata name (e.g. umeta0) whereas its value is indicated by $v_1$. The subsequent part presents the metadata values of all $j$ metadata of a resource, and first resource metadata name (e.g. rmeta0) and its value are represented by $m^r_1$ and $v_1$, respectively. The last part is a binary sequence with a `1' meaning `grant' and a `0' meaning `deny' for that operation. For example, $ \langle 1011 | 2021 | $\textit{30 49 5 26 63 129 3 42 $|$ 43 49 5 16 63 108 3 3 $|\langle$1 1 0 1}$\rangle\rangle$ is a sample authorization tuple of our dataset where 1011 and 2021 are the user and resource's unique number. The next eight numbers indicate the metadata values of a user, the following eight numbers represent resource's metadata values, and the final four binary digits signify the user has $op_1, op_2, op_4$ access to the resource. (In the example above, we also skip naming the metadata with the assumption that it could be inferred from the position of the metadata value.)  For simplicity, we assume the metadata values in our datasets are categorical, and each metadata value is an \textit{integer} representation of a category. We anticipate that our results will hold even in cases of metadata with \emph{real numbers}. 

\subsubsection{Dataset Visualization}
\input{graphics/fig-dataset-complexity-comparison-all}
We use t-SNE plots~\cite{van2008t-SNE} to visualize the samples in our datasets. A t-SNE plot discovers relationships in the data by identifying analogous clusters of data points with several features and projecting high dimensional features into a low dimensional feature space while retaining essential information.
We project each of our samples to a 2-dimensional feature and plot them. (Note that our datasets have varying numbers of features/metadata ranging from 16 to 26 in total.)
Each \textit{dot} in the plot (Figure~\ref{fig:complexDataset-tSNE}) represents an authorization tuple, where multiple tuples of the same color indicate that they have the same access permissions. For example, two tuples with only \textit{read} and \textit{write} access permissions will have the same color. Notably, a dataset with n operations will have tuples with $2^n$ different access combinations and plotted with $2^n$ distinct colors. For instance, the authorization tuples of a dataset with two operations (e.g., read and write) can be plotted with four different colors (tuples with the read access, write access, read and write access, and no access).

The position of a tuple in the plot is fixed according to the user and resource metadata values. For instance, two different users with the same metadata values may have access to a resource (or multiple resources with the same metadata values). Therefore, these tuples will have the same position regardless of their access permissions.
Figure~\ref{fig:complexDataset-tSNE}(a) depicts a dataset of $1650$ users and $320$ resources. The tuples (dots) take the position all over the plot according to their metadata values. This dataset has four different operations, and thereby, there are tuples with 16 distinct colors. 
However, we observe different tuples with the same access permission (same color) are grouped, and groups are isolated from one another, as shown in the figure with blue circles. Thus, a simple classifier can easily distinguish them without much difficulty (e.g., making one rule for each circle).
The access control states in real-world situations might be much more complicated~\cite{molloy2011adversaries}. Figure~\ref{fig:complexDataset-tSNE}(e) shows the visualization of a dataset from Amazon.
Even though this is not a complete access control scenario of the entire Amazon enterprise~\cite{cotrini2018Rhapsody}, samples with access (green dots) significantly overlap with other samples without access (red dots) which means tuples with very similar metadata values have entirely different accesses. A simple classifier would create too many rules to model such a dataset. 

\subsubsection{Introducing Complexity into Synthetic Datasets}
\label{sec:introRealWorldComplexity}
Informed by t-SNE visualization of the Amazon dataset, we seek to introduce complexity into our synthetic datatsets to closely reflect real-world situations. We observe that, in practice, the access privileges of users and resources with somewhat similar metadata could vary. It is also expected that not all the metadata of a user/resource would contribute equally to their permissions.
To reflect such scenarios in some of our datasets, we determine accesses considering \emph{all} the metadata values of user and resource but \emph{hide} a portion of metadata values from the policy miner (when dealing with mining methods) and model training phase (when dealing with ML methods). However, during rule generation, we ensure that the metadata that we will hide contributes to a lesser extent toward permission decisions by excluding them from being part of the \textit{constraint} of the rules. In a nutshell, hiding metadata attempts to simulate a scenario, where access decisions are made based on access control \emph{attributes} that are not fully informed by the entire metadata set. In a perfect world, access control attributes could succinctly capture the relevant metadata distributed across an organization. However, it is reasonable to hypothesize that this is not a practical assumption.

Figure~\ref{fig:complexDataset-tSNE}(b) represents a dataset with 11 user and 11 resource metadata. The authorization tuples were created considering all the metadata. To simulate a similar situation while visualizing tuples and understand how it looks from a policy mining (or classification) perspective, we make only the first eight user and the first eight resource metadata values available for visualization. As shown, many dots of other colors now start to mix, indicating two tuples with similar user-resource metadata values may have very different accesses.
Indeed, such proximity of the tuples is challenging for clustering or rule creation. The more metadata we hide, the more complicated the dataset for policy mining and ML approaches.

\input{graphics/fig-dataset-all-datasets}
\input{tables/tab-dataset}

We still notice a few portions in the plot where the same colored dots are clustered together and separable, as shown with red circles in Figure~\ref{fig:complexDataset-tSNE}(b). This is because the dataset generation algorithm~\cite{xu2014mining} creates the metadata values based on a distribution, where the value range (i.e., number of values for each metadata) is sparse. For example, there are around one hundred different values for specific metadata (e.g., \emph{department}) in a dataset with hundred users. One can easily cluster all the users into hundreds of groups based only on the department. However, there might be the case where metadata values are required to be chosen from a set of a limited number of values (say, ten departments for hundred users). Evidently, it is harder to cluster one hundred users into ten groups than a hundred. To reflect this, for each metadata, we define a fixed and smaller set of values (6 to 20 unique values) following the same distribution used by Xu et al.~\cite{xu2014mining}. We choose each metadata value from the corresponding list during user/resource creation and metadata value assignment.
This strategy creates datasets with significantly overlapped samples as depicted in Figure~\ref{fig:complexDataset-tSNE}(c).
We also extended the number of users and resources to simulate a larger organization, which adds more overlaps among samples and higher complexity to the dataset, as shown in Figure~\ref{fig:complexDataset-tSNE}(d).

Finally, we synthesized eight different datasets (datasets \#3-\#10 in Table \ref{tab:ourDataset}) used for \(\DLBACALPHA\) experimentation and evaluation, with varying numbers of users, resources, user and resource metadata, and authorization tuples, each reflecting a varying degree of complexity. We use the following naming convention for our synthetic datasets, as listed in Table~\ref{tab:ourDataset}:
$u\langle$approx. number of users$\rangle-r\langle$approx. number of resources$\rangle-auth\langle$approx. number of authorization tuples$\rangle$.
We use `u', `r', and `auth' to indicate users, resources, and authorization tuples, respectively. We also visualize all the synthetic datasets in Figure~\ref{fig:allDatasets-tSNE}(a-h) using t-SNE plots. As illustrated in the Figure, each plot has dots with one of 16 different colors (for four operations), and those dots mix extensively.


\subsection{Real-world Dataset}
\label{sec:realworldDataset}
Amazon published two datasets that contain access control information which is widely used in access control research~\cite{cotrini2018Rhapsody, cappelletti2019quality, karimi2021automatic}. We name these datasets as \(\AMAZONKAGGLE\) and \(\AMAZONUCI\), and list them in the Table~\ref{tab:ourDataset} (1-2).
The \(\AMAZONKAGGLE\) dataset was released in Kaggle~\cite{AmazonKaggle2013} (a platform for predictive modeling competitions) as a challenge to the community to build a machine learning model to determine the employees' accesses. The dataset holds historical access data where employees were manually allowed or denied access to resources over time. The dataset has about nine thousand users and seven thousand resources with over 32K authorizations tuples. Each tuple specifies eight user metadata that depicts a user's properties, a resource id to identify the resource, and a binary flag to indicate whether the user has access to the resource or not. However, the dataset is highly imbalanced, and about 93\% of the tuples are with grant access. We visualize this dataset in Figure~\ref{fig:allDatasets-tSNE}(i). As we see, there are dots from two colors where green and red correspond to tuples with grant and deny access, each. Notably, a significant number of dots are from grant accesses.

The \(\AMAZONUCI\) dataset was provided by Amazon in the UCI machine learning repository~\cite{AmazonUCI2011}. This dataset contains access information of more than 36,000 users and 27,000 permissions. For any permission, less than 10\% of all users have requested access. The dataset is widely used in access control researches, and in most of the cases, the experiments are confined to only 5 to 8 most requested permissions~\cite{cappelletti2019quality, cotrini2018Rhapsody, karimi2021automatic}. Likewise, we took the seven most requested permissions, and for each permission, we list users who have access to the selected permissions. However, this dataset is also imbalanced, around 75\% tuples with the deny access. In addition, the dataset is not ABAC in nature, and there are some tuples in the dataset where users with identical attribute values do not have the same access permissions. Because of this, policy mining or classification approaches may suffer while clustering the users for different permissions. 
We visualize this dataset in Figure~\ref{fig:allDatasets-tSNE}(j).

\subsection{Neural Network Architecture and Training}
\label{sec:organizingTrainingData}

\begin{figure}[t]
\centering
\includegraphics[width=\linewidth] {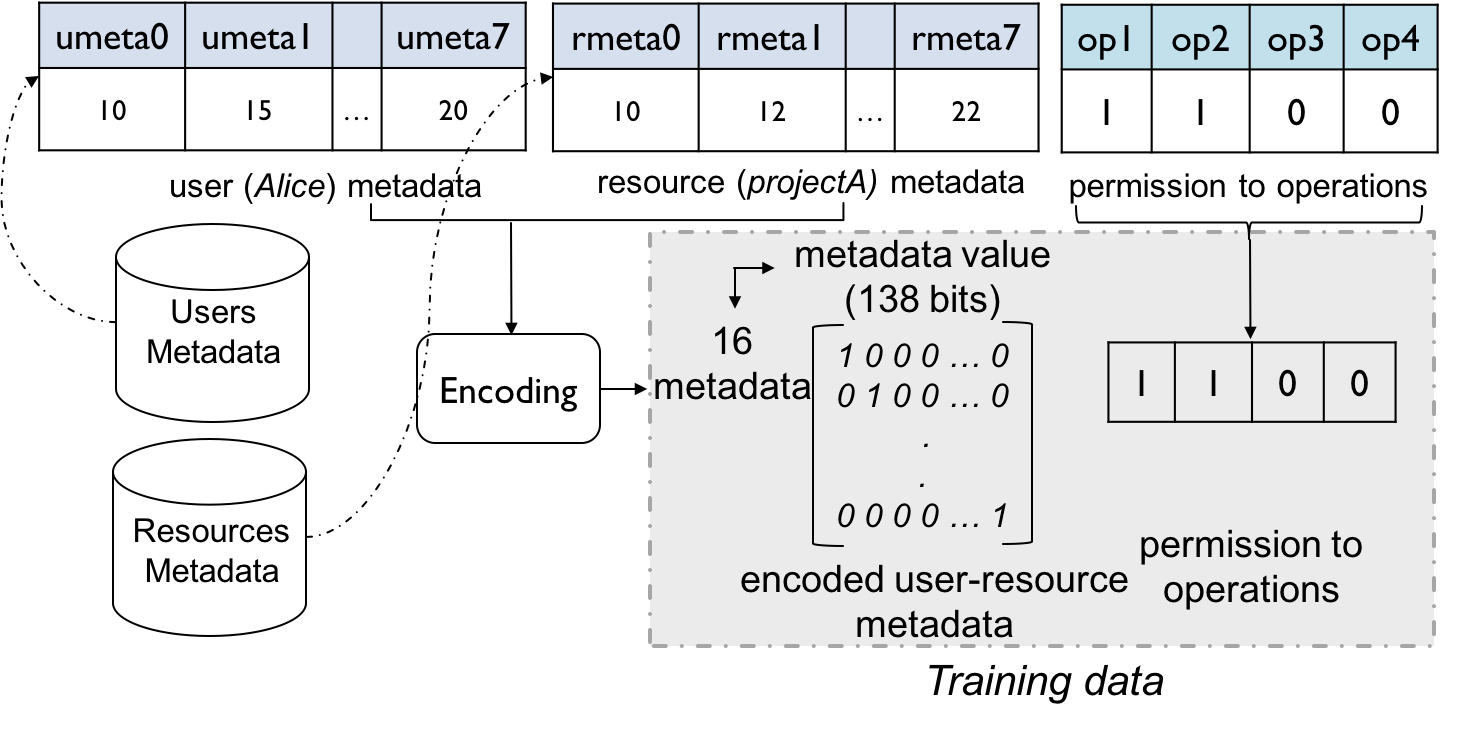}
\caption{Preparing Training Data for \(\DLBACALPHA\).}
\label{fig:organizingTrainingDataForDLBACalpha}
\vspace{-2ex}
\end{figure}

For \(\DLBACALPHA\), the deep neural network takes user/resource metadata values as input. It includes a
classification layer with the number of neurons equal to the number of operations, where each neuron outputs the \emph{probability} of granting the permission for a related operation, \textit{op}. 
Given a feature vector $x$ of the user and resource metadata, the neural network can be defined as a prediction function $f$ such that $\hat{y} = f(x)$, where $\hat{y}$ is the predicted label or permission (grant (1) or deny (0)) of $op$, obtained from comparing the probability of granting the permission at the output of the network with a \emph{threshold}. 
Note that for all \(\DLBACALPHA\) experimentation, we consider a threshold of 0.5.
To train the neural network $f$ (i.e. determine the network's weights), a set of training authorization tuples $\mathbb X$ of size $N$ is collected, where $(x_{i}, y_{i})$ denotes the $i$-th sample in $\mathbb X$, where $x_{i}$ is the feature vector of the user and resource metadata and $y_{i}$ is the corresponding $op$ or the target label. 
As discussed in Section~\ref{sec:introRealWorldComplexity}, for some synthetic datasets, we hide a portion of metadata from the policy mining algorithm to mimic some complex situations. We apply this for the datasets having more than eight user metadata and eight resource metadata. In such a case, $x_{i}$ represents the feature vector of the first \emph{eight user metadata} and the first \emph{eight resource metadata}. So, for example, for a dataset with 13 user-resource metadata, we hide five metadata from the user metadata and five from the resource metadata.

Since the metadata values in our dataset are categorical, we map them to binary values by utilizing encoding. We encode a tuple's user/resource metadata value using \textit{one-hot encoding}~\cite{hancock2020survey} to transform the categorical values into a two-dimensional binary array. The row in the array represents metadata, and the column holds the encoded binary representation of the corresponding metadata value.
(Amazon dataset's metadata values are too sparse, and we use \emph{binary encoding} for them considering its memory efficiency in such cases~\cite{seger2018investigation}).
As each operation is binary, we apply them directly as the target labels without any processing. Figure~\ref{fig:organizingTrainingDataForDLBACalpha} illustrates the overall training data preparation for \(\DLBACALPHA\). As illustrated, we encode the metadata values of a user $Alice$ and a resource $projectA$ and apply permissions to the associated operations without further processing.
We train \(\DLBACALPHA\) based on the training data, 
and this \textit{trained} \(\DLBACALPHA\) is used in \textit{Access Control Decision Engine} (discussed below) to produce access decisions for test data.


\begin{figure}[t]
\centering
\includegraphics[width= 0.8\linewidth, scale = 0.9] {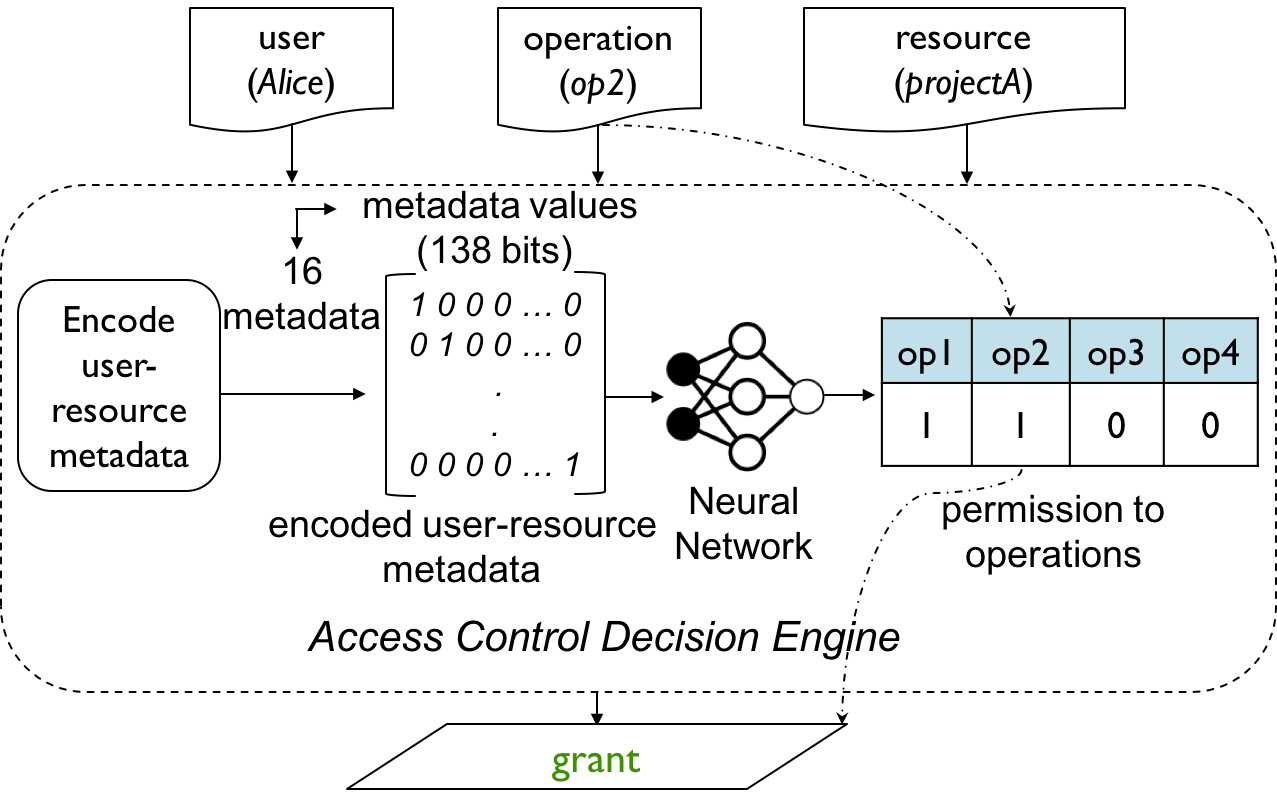}
\caption{Decision Making Process in \(\DLBACALPHA\).}
\label{fig:decisionMakingProcessDLBACalpha}
\vspace{-1ex}
\end{figure}

\subsection{Decision Making Process in \(\DLBACALPHA\)}
\label{sec:decisionProcessDLBACalpha}

\textit{Access Control Decision Engine (Decision Engine)} is a DLBAC component responsible for receiving and authorizing any access request. In \(\DLBACALPHA\), the \textit{Decision Engine} (Figure~\ref{fig:decisionMakingProcessDLBACalpha}) takes three inputs (\textit{user, resource, and operation}).
The \textit{Decision Engine} retrieves the user and resource metadata from the internal databases and then encodes them to obtain corresponding binary representation.
The encoded input is fed into the neural network to predict the corresponding request's access permission. The network outputs access information for all the operations. The decision engine then determines the actual access authorization based on the requested access and the network's output.
For the specific example in Figure~\ref{fig:decisionMakingProcessDLBACalpha}, user \emph{Alice} wants \emph{op2} access on \emph{projectA} resource. The output of the neural network for the \emph{op2} is $1$, which indicates that \emph{Alice} has \emph{op2} access on \emph{projectA}. Thus, the \textit{Decision Engine} authorizes this request.

%% file: graphics/fig-dataset-complexity-comparison-all.tex
\begin{figure*}[t]
\vspace{-2ex}
\centering
	\includegraphics[width=\textwidth]{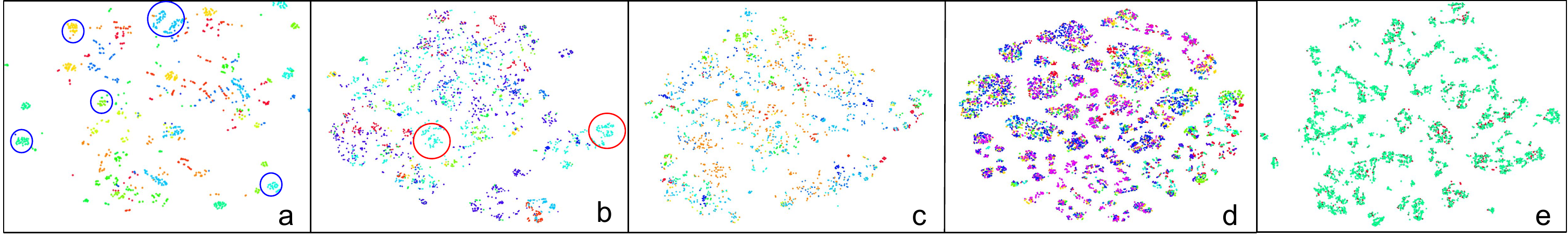}
	\caption{Comparing Complexity of Datasets. (a) A dataset with 1650 users, 320 resources, eight user/resource metadata, (b) A dataset with 1000 users, 639 resources, 11 user/resource metadata, (c) A dataset with 800 users, 656 resources, 11 user/resource metadata, (d) 4500 users, 4500 resources, 11 user/resource metadata, (e) An Amazon dataset~\cite{AmazonKaggle2013} (The dataset has more samples with `grant' access. Therefore, for better visualization, we considered all the tuples with `deny' access permission and randomly selected a similar number of tuples with `grant' access permission).}
	\label{fig:complexDataset-tSNE}
	\vspace{-2ex}
\end{figure*}


%% file: graphics/fig-dataset-all-datasets.tex
\begin{figure*}[t]
\vspace{-2ex}
\centering
	\includegraphics[width=\textwidth]{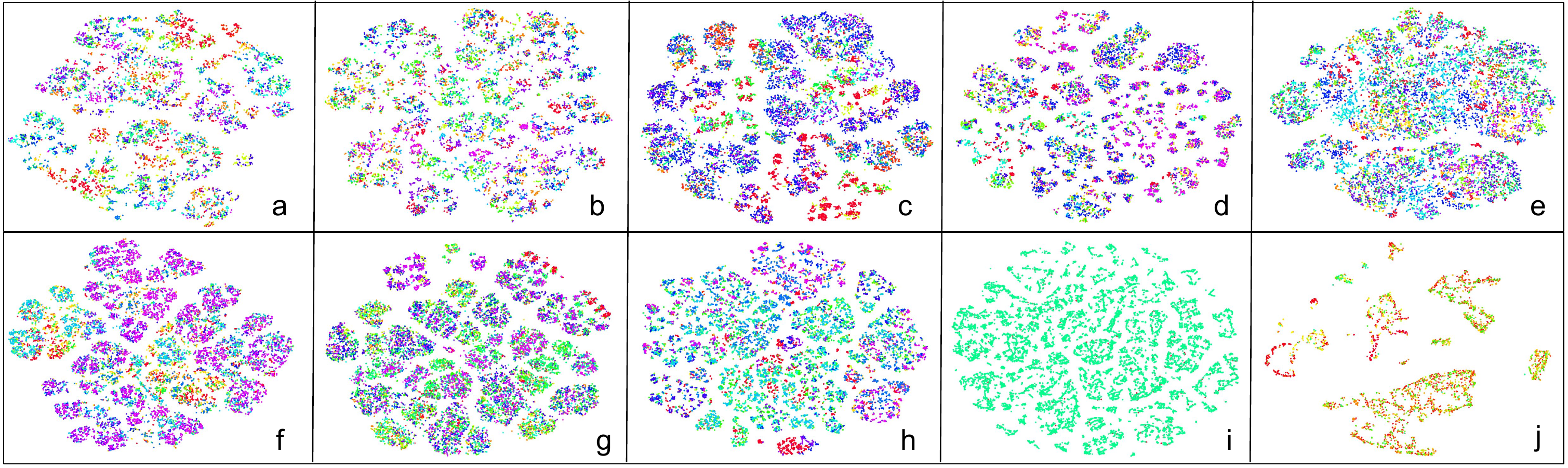}
	\caption{t-SNE Visualization of Synthetic and Real-world Datasets from Table~\ref{tab:ourDataset}. Figures a-h corresponds to synthetic datasets \#3$-$\#10 and i-j corresponds to real-world datasets \#1$-$\#2 in Table~\ref{tab:ourDataset}, respectively.}
	\label{fig:allDatasets-tSNE}
	\vspace{-2ex}
\end{figure*}

%% file: tables/tab-dataset.tex
\begin{table}[t]
\vspace{-2ex}
\centering
\normalsize
  \caption{List of Datasets.}
  \label{tab:ourDataset}
  \resizebox{\linewidth}{!}{%
  \begin{tabular}{cccccccc}
    \hline
    \#
    & Dataset
    & Type
    & Users
    & \multicolumn{1}{p{1.1cm}}{\centering User\\Metadata}
    & Resources
    & \multicolumn{1}{p{1.1cm}}{\centering Resource\\Metadata}
    & \multicolumn{1}{p{0.8cm}}{\centering Authorization\\Tuples}\\\hline
    1 & \(\AMAZONKAGGLE\) & Real-world & 9560 & 8 & 7517 & 0 & 32769\\
    2 & \(\AMAZONUCI\) & Real-world & 4224 & 11 & 7 & 0 & 4224\\
    3 & \(\SMALLA\) & Synthetic & 4500 & 8 & 4500 & 8 & 10964\\
    4 & \(\SMALLB\) & Synthetic & 5250 & 8 & 5250 & 8 & 12690\\
    5 & \(\MEDIUMA\) & Synthetic & 5250 & 10 & 5250 & 10 & 19535\\
    6 & \(\MEDIUMB\) & Synthetic & 4500 & 11 & 4500 & 11 & 20979\\
    7 & \(\MEDIUMC\) & Synthetic & 4500 & 11 & 7194 & 11 & 20033\\
    8 & \(\MEDIUMD\) & Synthetic & 4500 & 13 & 4500 & 13 & 22583\\
    9 & \(\LARGEA\) & Synthetic & 4500 & 13 & 6738 & 13 & 28751\\
    10 & \(\LARGEB\) & Synthetic & 6000 & 10 & 6000 & 10 & 32557\\
    \hline
  \end{tabular}
  }
  \vspace{-3ex}
\end{table}

%% file: contents/evaluation.tex
\section{Evaluation} \label{sec:evaluation}
In this section, we experimentally evaluate the performance of \(\DLBACALPHA\) using both synthetic and real-world datasets. 


\subsection{Evaluation Methodology}
\label{sec:evaluationMethodology}
We experiment 
and evaluate the performance for all the datasets listed in Table~\ref{tab:ourDataset}. We consider each dataset to represent an organization with its own unique characteristics.
We split each dataset into training (80\%) and testing (20\%) sets. As the test dataset is entirely \textit{unseen} during training, the evaluation shall adequately measure the generalization of any method. 

\textbf{Instances of \(\DLBACALPHA\).} DLBAC is agnostic to deep neural network architecture, and we will show that the deep learning-based model's performance is consistent across datasets. For demonstration, we implement three instances of \(\DLBACALPHA\) using three distinct deep neural network architectures including ResNet~\cite{he2016deep}, DenseNet~\cite{huang2017densely}, and Xception~\cite{chollet2017xception}, and name them as \(\DLBACALPHAR\), \(\DLBACALPHAD\), and \(\DLBACALPHAX\), respectively. For \(\DLBACALPHAR\), we use the ResNet network with depth 8 for the first four datasets in Table~\ref{tab:ourDataset} whereas, for the rest of the datasets, we use a ResNet with depth 50. For \(\DLBACALPHAD\), we use the DenseNet architecture with [6,12,24,16] layers in the four dense blocks. We adopt the source code from the Keras application for all the model architectures implementation.\footnote{\url{https://github.com/keras-team/keras-applications}}

The \(\DLBACALPHA\) instances were developed in Python using Keras library with a TensorFlow backend and trained on Google Colab (a 12GB NVIDIA Tesla K80 GPU).
We apply Adam optimizer with an initial learning rate of 0.001, scheduled to reduce the learning rate by dividing by ten after every 10 epochs. The epoch and batch size was chosen as 60 and 16, respectively, with an early stop after 5 consecutive epochs without any performance improvement. As the \(\DLBACALPHA\) outputs an access probability between `0' and `1' for each operation, we use binary cross-entropy loss. We have created a repository in GitHub consisting of all the datasets, source code, and trained networks.\footnote{\url{https://github.com/dlbac/DlbacAlpha}}

\textbf{Machine learning (ML) algorithms.} 
We compare the performance of \(\DLBACALPHA\) instances with classical machine learning approaches such as Support Vector Machine (SVM)~\cite{cortes1995support} and Random Forests (RF)~\cite{breiman2001random}. We also compare with Multi-Layer Perceptron (MLP)~\cite{schmidhuber2015deep} with four hidden layers (a shallow network) to evaluate how significant the performance difference is between a deep and a non-deep neural network.
We use the \textit{SVC} and \textit{RandomForestClassifier} class of the Python scikit-learn library~\cite{scikit-learn} for SVM and RF implementation, respectively, with their default configurations. We implement MLP using Keras library.

\textbf{Policy mining algorithms.} 
There is no other existing \textit{deep learning-based} access control approach to the best of our knowledge, so a direct comparison of our work results is not currently possible. Therefore, we compare \(\DLBACALPHA\) with ABAC policy mining algorithms being one of the flexible and generalized access control approaches.
We compare the performance of \(\DLBACALPHA\) instances with the following policy mining algorithms. While a few other works exist as discussed in our related work, a key decision factor in selecting these works was our ability to readily access their source codes and our ability to clearly understand, modify/tweak as needed and compile them.
\begin{enumerate}
\item The policy mining algorithm proposed by Xu and Stoller~\cite{xu2014mining}, which we refer to as XuStoller.
\item Rhapsody~\cite{cotrini2018Rhapsody}, a policy mining algorithm built upon an ML algorithm for subgroup discovery named APRIORI-SD~\cite{kavvsek2006apriori-sd}. Rhapsody performance has a direct correlation with multiple parameters. We trial with different parameter values and selected the policy with the highest F1 score while maintaining an FPR below 0.05.
\item EPDE-ML~\cite{liu2021efficient}, a permission decision engine scheme based on ML where a trained RF model makes the access control decision.
\end{enumerate}

\textbf{Evaluation metrics.}
For ML algorithms, we compute the F1 score and compare the performance with \(\DLBACALPHA\) instances and show that deep learning based algorithms generally perform better than traditional ML and MLP techniques. For an extensive comparison against policy mining algorithms, we compute the F1 score, False Positive Rate (FPR), True Positive Rate (TPR), and Precision. We consider the standard definitions~\cite{liu2021efficient, cotrini2018Rhapsody} of these evaluation metrics. Policies (or models) with a higher F1 score lead to better generalization. They can make more accurate access control decisions on users and resources with attributes not explicitly seen during the mining (or training) process. Also, the higher TPR and Precision are better as these scores indicate how accurately and efficiently the policies (or models) can grant access. On the contrary, the policies (or models) with a lower FPR are better as they are less likely to give access to requests, those which should be denied according to the ground truth access control policy.

\subsection{Results}
\subsubsection{Performance comparison with ML algorithms}

\begin{figure}[t]
\vspace{-2ex}
\centering
\includegraphics[width=\linewidth]
{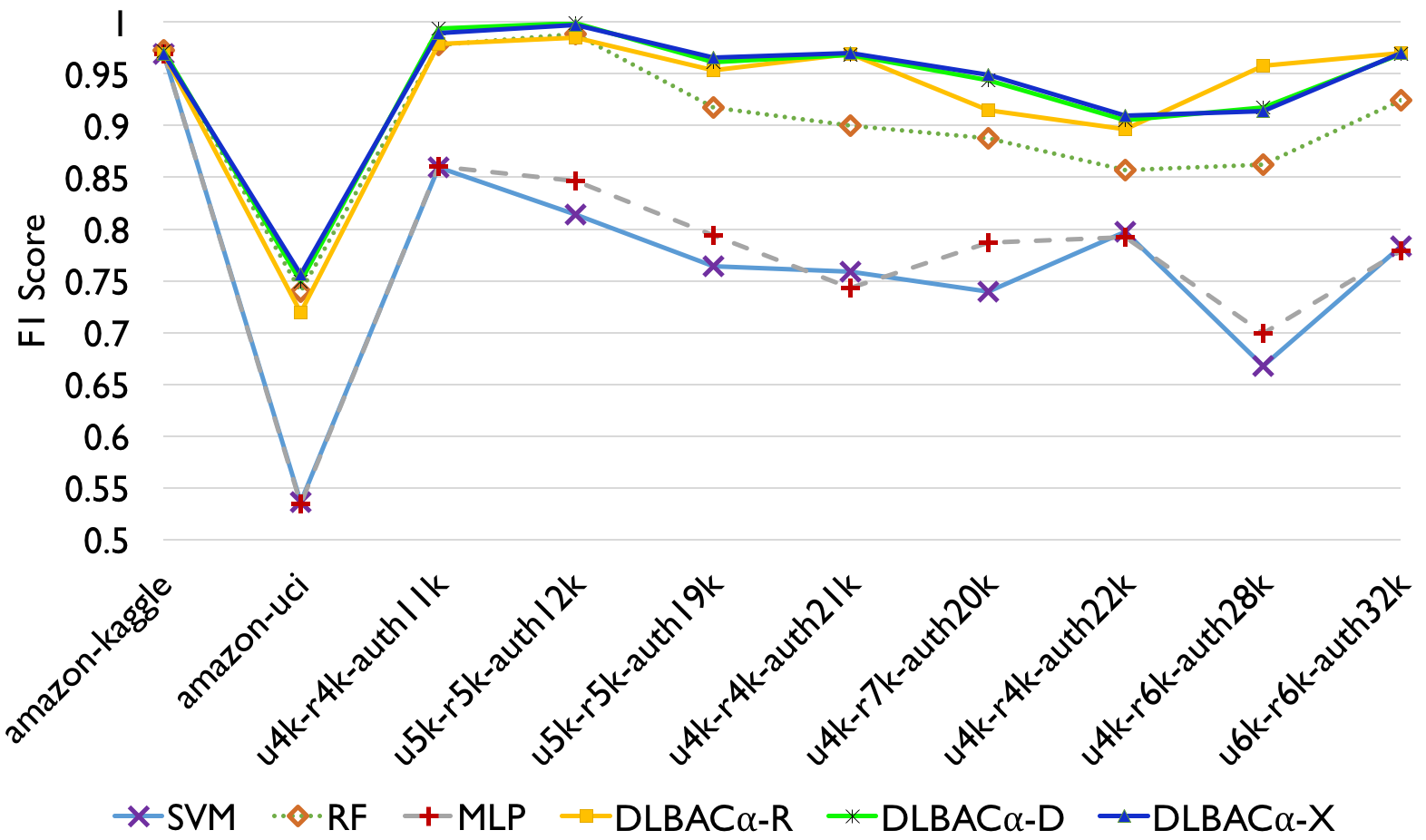}
\caption{F1 Score Comparison: ML Algorithms vs. \(\DLBACALPHA\) Instances.}
\label{fig:determiningEffectiveMLTechnique}
\vspace{-2ex}
\end{figure}

Figure~\ref{fig:determiningEffectiveMLTechnique} illustrates the overall performance of all ML approaches and \(\DLBACALPHA\) instances for each dataset with respect to F1 score. The performance of all the algorithms is consistent and better for the \(\AMAZONKAGGLE\) dataset, but it is not the case for \(\AMAZONUCI\) dataset. In this case, SVM and MLP performed significantly lower, and other approaches including \(\DLBACALPHA\) instances could not achieve high performance. Such a result is expected due to the inconsistency in the access permissions in that dataset. The dataset contains tuples where users with identical attribute values do not have the same access permissions, as discussed in Section~\ref{sec:realworldDataset}.
The advantage of the DLBAC approach is more evident if the dataset is processed correctly. We show this using our synthetic datasets. For synthetic datasets, \(\DLBACALPHA\) instances achieved the highest F1 score, while SVM and MLP performed the worst. 
Also, the instances of \(\DLBACALPHA\)'s improvements over RF are significant (for p-value $<$ 0.05; paired T-test (not shown here)) for all the synthetic datasets except \(\SMALLB\) dataset.
The performance advantage of \(\DLBACALPHA\) instances is particularly pronounced in synthetic datasets with a large number of authorization tuples, where \(\DLBACALPHA\) instances report 0.03 to 0.09 improvements over RF as shown in the figure. Notably, the performances of all algorithms vary with the complexity of the datasets. However, \(\DLBACALPHA\) instances show the lowest variation in its performance across the datasets, implying that \(\DLBACALPHA\) is most robust against changes in data characteristics such as number of hidden metadata, number of users, resources, and authorization tuples. Except for the \(\LARGEA\) dataset, all algorithms' performances drop with the increase of hidden metadata (e.g., the \(\MEDIUMD\) dataset with 13 metadata where we hide 5 of the metadata from the feature vector as discussed in Section~\ref{sec:organizingTrainingData}), suggesting that an increase in data complexity generally impacts performance. Overall, the experimental results indicate that \(\DLBACALPHA\) is more effective and robust than classical ML approaches, including MLP, for making accurate access decisions.

While the performance advantages of \(\DLBACALPHA\) are not apparent in the Amazon datasets, we emphasize that those datasets are not reflective of the access state complexity of the entire organization but that of a small portion of the company. This is one of the reasons why we synthesized additional datasets for our experimentation.

\subsubsection{Performance comparison with policy mining algorithms}

\begin{figure}[t]
\vspace{-2ex}
\centering
\includegraphics[width=\linewidth]
{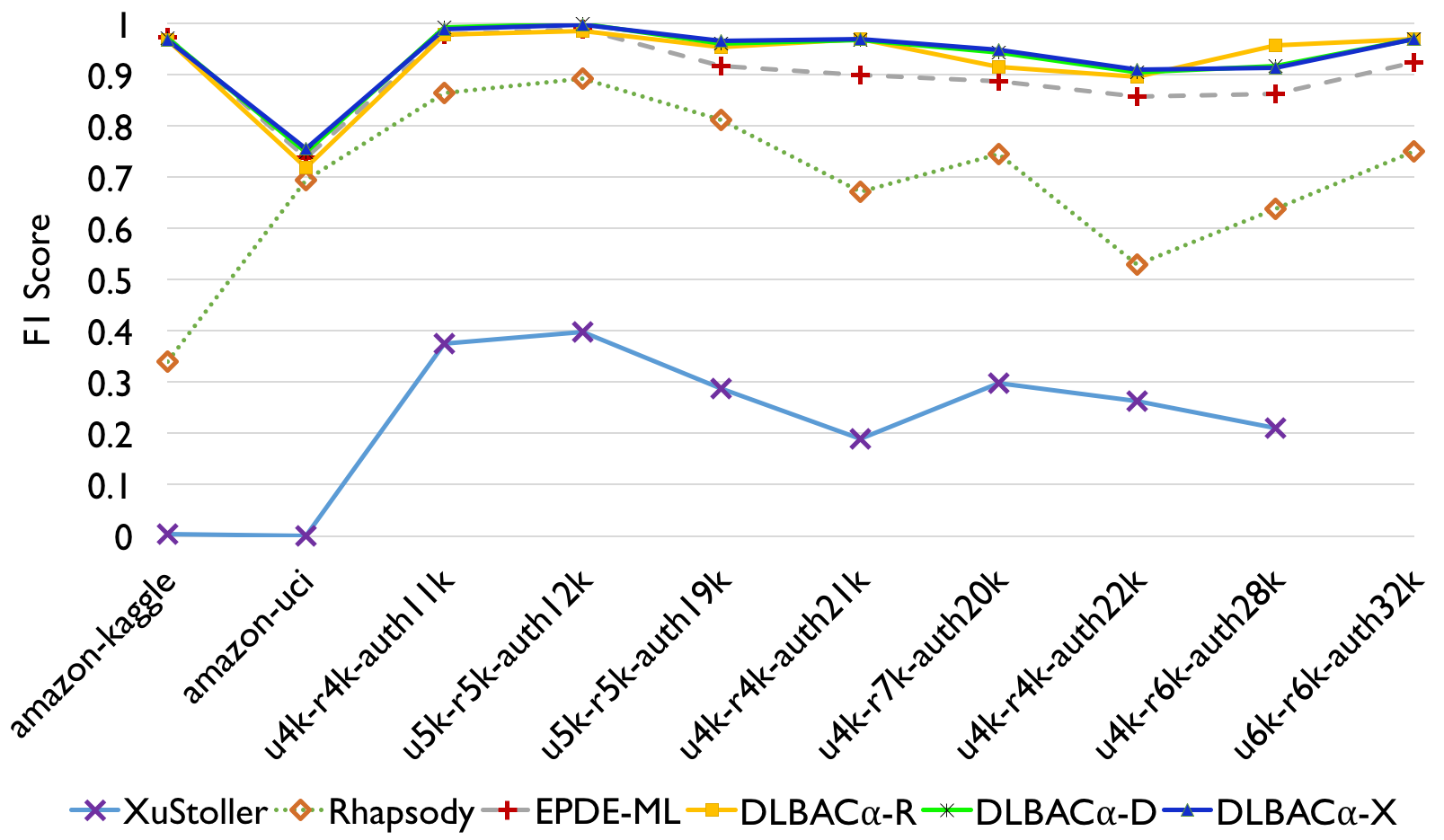}
\caption{F1 Score Comparison: Policy Mining Algorithms vs. \(\DLBACALPHA\) Instances.}
\label{fig:f1compareDLBACalphaPolicyMining}
\vspace{-2ex}
\end{figure}

\begin{figure}[t]
\vspace{-1ex}
\centering
\includegraphics[width=\linewidth]
{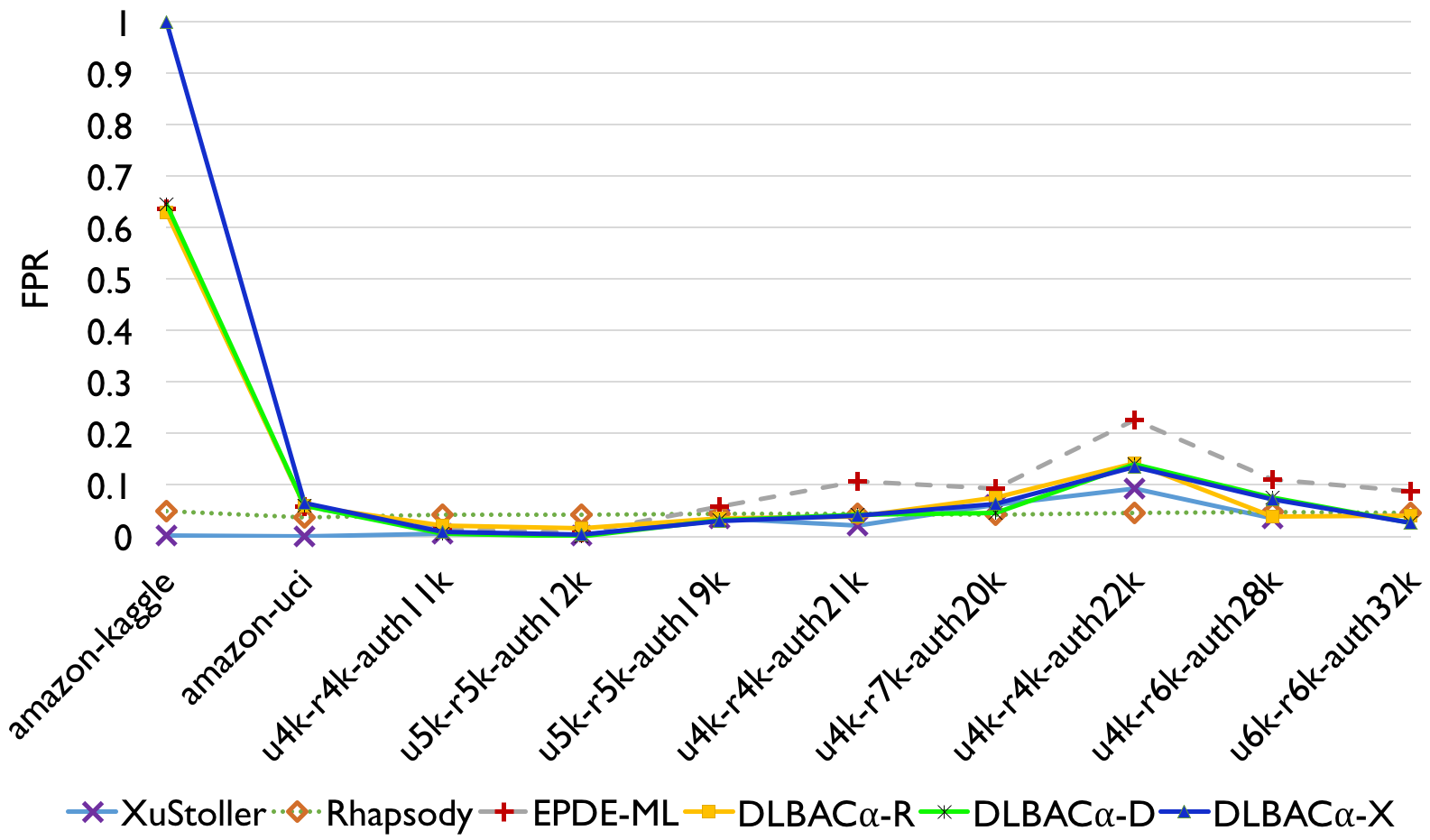}
\caption{FPR Comparison: Policy Mining Algorithms vs. \(\DLBACALPHA\) Instances.}
\label{fig:fprCompareDLBACalphaPolicyMining}
\vspace{-2ex}
\end{figure}

\begin{figure}[t]
\vspace{-1ex}
\centering
\includegraphics[width=\linewidth]
{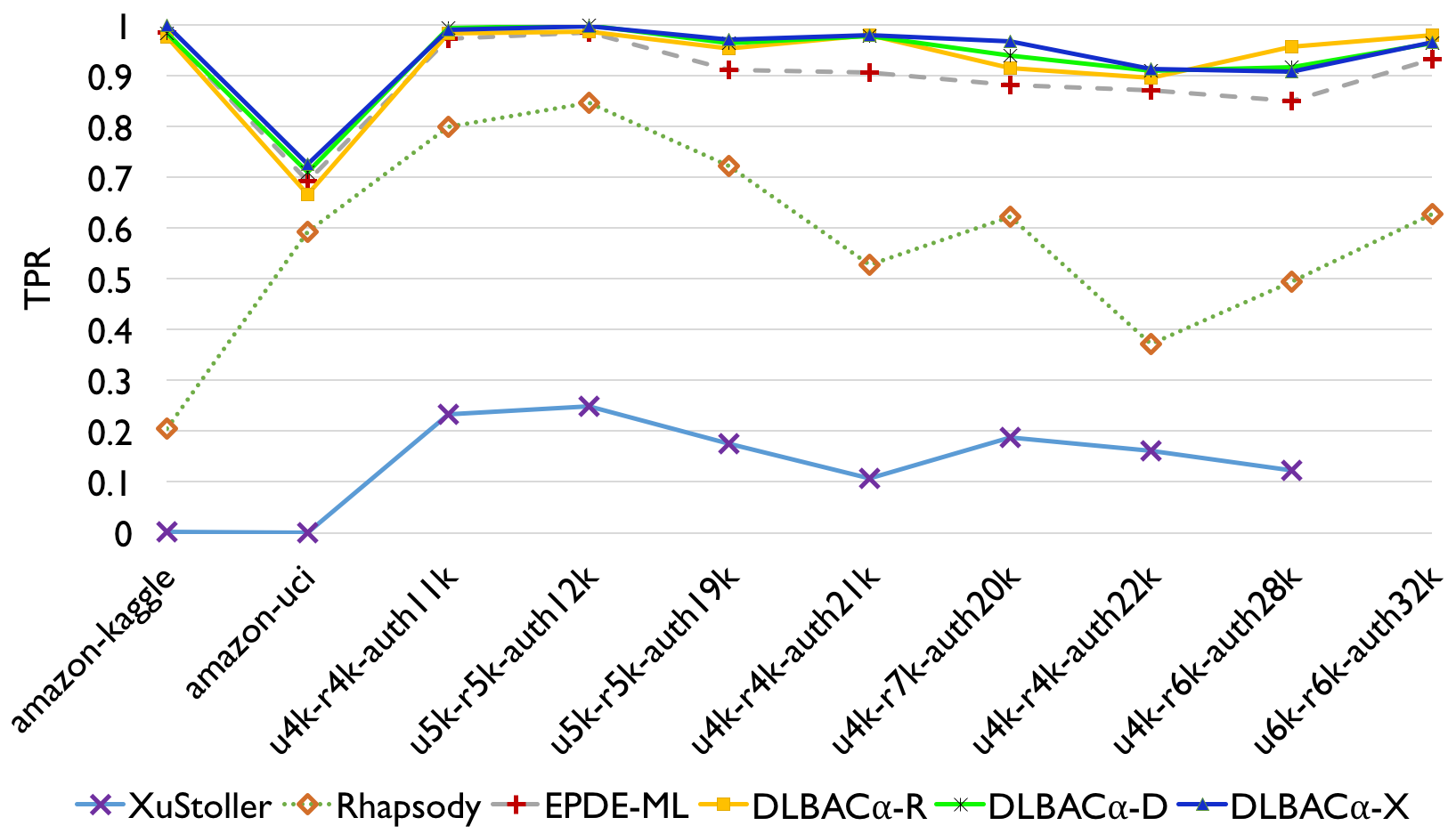}
\caption{TPR Comparison: Policy Mining Algorithms vs. \(\DLBACALPHA\) Instances.}
\label{fig:tprCompareDLBACalphaPolicyMining}
\vspace{-2ex}
\end{figure}

\begin{figure}[t]
\vspace{-2ex}
\centering
\includegraphics[width=\linewidth]
{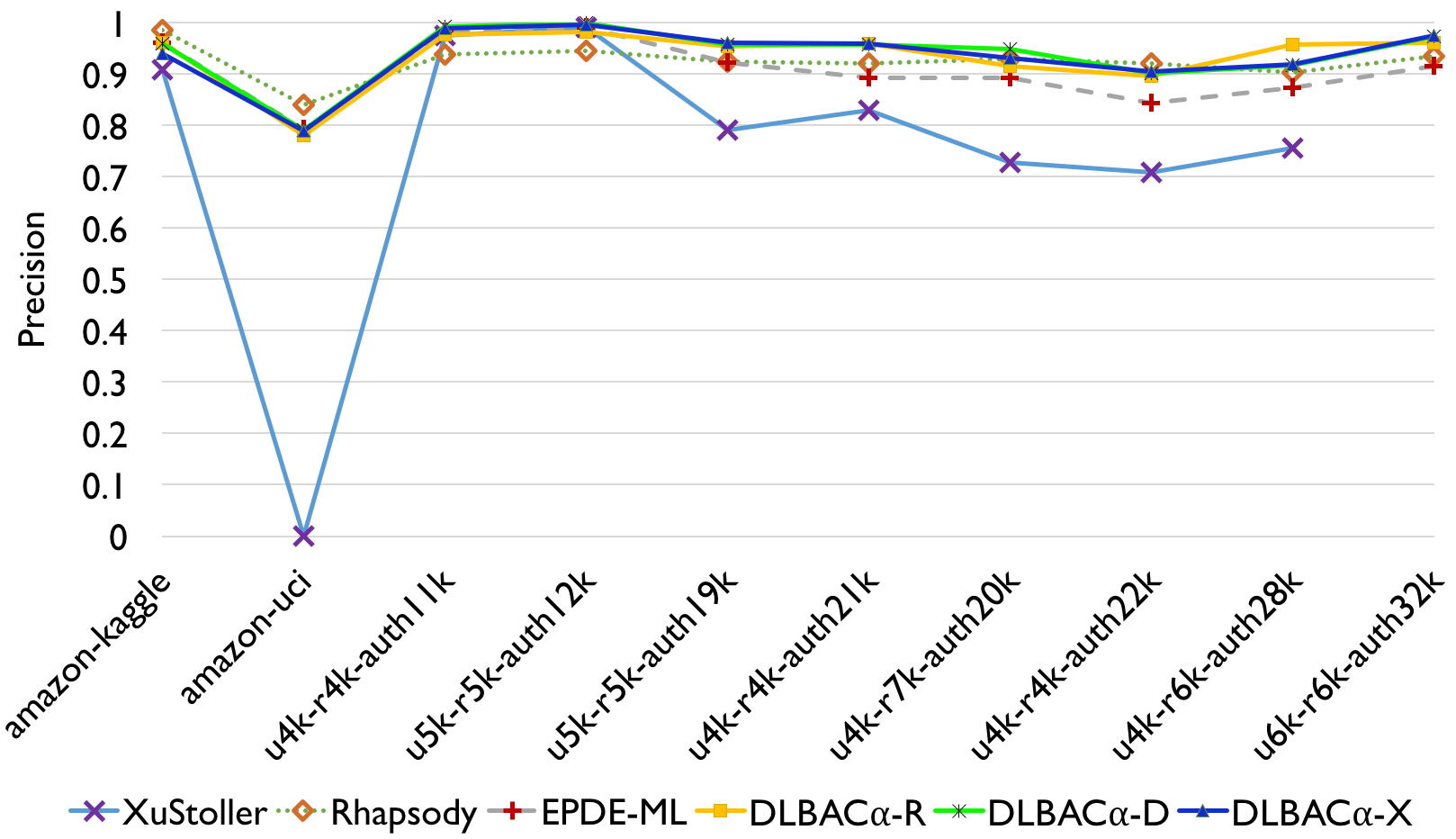}
\caption{Precision Comparison: Policy Mining Algorithms vs. \(\DLBACALPHA\) Instances.}
\label{fig:precisionCompareDLBACalphaPolicyMining}
\vspace{-2ex}
\end{figure}

Figures~\ref{fig:f1compareDLBACalphaPolicyMining},~\ref{fig:fprCompareDLBACalphaPolicyMining},~\ref{fig:tprCompareDLBACalphaPolicyMining}, and ~\ref{fig:precisionCompareDLBACalphaPolicyMining} compare the F1 score, FPR, TPR, and Precision, respectively of policy mining algorithms and \(\DLBACALPHA\) instances. We could not experiment XuStoller algorithm for the largest synthetic dataset (\(\LARGEB\)) as it took a very long runtime without any output. We can make the following observations based on all these experimental results.
\begin{itemize}
    \item \textit{A deep learning based approach can make more accurate access control decisions and generalize better.} The F1 score of EPDE-ML and \(\DLBACALPHA\) instances are significantly better than the rule-based approaches such as XuStoller and Rhapsody. That signifies, in general, machine learning-based approaches can make better generalization and accurate access control decisions. As we see in Figure~\ref{fig:f1compareDLBACalphaPolicyMining}, the performance improvements of \(\DLBACALPHA\) instances over EPDE-ML, which is statistically significant for most datasets, suggest that deep learning based algorithms make the more accurate decision and have even better generalization capability than classical ML-based policy mining approaches.
    
    \item \textit{A deep learning based approach can properly balance both over-provision and under-provision.} XuStoller and Rhapsody achieved the best FPR as shown in Figure~\ref{fig:fprCompareDLBACalphaPolicyMining}, indicating they are unlikely to give access to requests that should be denied according to the actual access control policy. However, that is not the case for denying access. As we see in Figure~\ref{fig:tprCompareDLBACalphaPolicyMining}, the TPR for Rhapsody is between 0.2 to 0.85, and for XuStoller, it is from 0 to 0.25. Such a lower TPR indicates that these algorithms are pretty inefficient while denying access (under-provisioned) even though the requests deserve grant access according to the actual access control policy. 
    On the other hand, the EPDE-ML performed lowest in terms of FPR (over-provisioned) across synthetic datasets ranging from 0.06 to 0.23. Their average TPR and Precision are below 0.9.
    Such higher FPR and comparably lower TPR and Precision suggest that EPDE-ML could not achieve desirable performance in terms of over-provision and under-provision. The \(\DLBACALPHA\) instances obtained a much higher TPR and Precision as illustrated in Figure~\ref{fig:tprCompareDLBACalphaPolicyMining} and~\ref{fig:precisionCompareDLBACalphaPolicyMining}. Also, the  \(\DLBACALPHA\) instances reached an FPR which is comparable to XuStoller and Rhapsody, as shown in Figure~\ref{fig:fprCompareDLBACalphaPolicyMining}. This suggests that deep learning based approach can balance better between over- and under-provisioning. 
    
    \item \textit{An imbalanced dataset may affect some performance that could be calibrated.} The FPR result of \(\DLBACALPHA\) instances for \(\AMAZONKAGGLE\) dataset is high, and the TPR for \(\AMAZONUCI\) dataset is relatively low, arising due to the characteristics of the dataset~\cite{weng2008new}. As discussed in Section~\ref{sec:realworldDataset}, these datasets are imbalanced, one has unreasonably more samples from the grant class, and the other has more from the deny class.
    We argue that this is a typical machine learning problem, and EPDE-ML has a similar performance for these datasets. For balanced datasets, the FPR and TPR of \(\DLBACALPHA\) instances are consistent, and FPR is below 0.05 for most of the datasets while TPR is above 0.95, as demonstrated in Figure~\ref{fig:fprCompareDLBACalphaPolicyMining} and~\ref{fig:tprCompareDLBACalphaPolicyMining}. Evidently, these metrics could be calibrated based on the tolerance to over vs. under-provisioning of an application context. Specifically, one could favor a particular metric in \(\DLBACALPHA\) by modifying the loss function to increase the weight of the minority class~\cite{ho2019real}, 
    adjusting the threshold for granting permissions as described in Section~\ref{sec:organizingTrainingData}, etc. We note that an improvement in one of the metrics will likely negatively impact one or more of the others and that every work is prone to this issue.
\end{itemize}

Overall, the \(\DLBACALPHA\) instances achieved better or comparable performance for all the metrics across datasets, suggesting that deep learning based approaches generalized better and made more accurate access control decisions than rule based and classical ML based approaches.
These results demonstrate the effectiveness of using DLBAC as an access control system.

%% file: contents/interpretability.tex
\section{Understanding DLBAC Decisions} \label{sec:understandability}

As the core of a DLBAC system is a \textit{neural network}, a major challenge is providing insights into why and how DLBAC makes certain decisions. That is, explainability is a key challenge for DLBAC.
For instance, in Figure~\ref{fig:decisionMakingProcessDLBACalpha}, the \textit{Decision Engine} received a request that user \textit{Alice} wishes \textit{op2} access on \textit{projectA} resource. 
Based on the result of the \textit{neural network}, the decision engine approved the request. However, it is not quite obvious why the \textit{neural network} made that prediction for this request. Such a justification is generally straightforward in traditional access control systems as the decisions are made based on written policies. But, it is challenging for DLBAC due to the \textit{black-box} nature of a  neural network~\cite{shrikumar2017learning}. 
As the decisions are made based on provided \textit{user/resource metadata}, it is essential to understand why a decision is made and which metadata influenced that decision.
Many techniques have been introduced to help gain insights into a neural network's internal details. In this section, we investigate two state-of-the-art approaches for this purpose: \textit{Integrated Gradients}~\cite{sundararajan2017axiomatic} and \emph{Knowledge Transferring}~\cite{fukui2019distilling}. We experiment on \(\DLBACALPHAR\) instance (as introduced in Section~\ref{sec:evaluationMethodology}), and for brevity, we refer it to \(\DLBACALPHA\) in the following discussion. 
The related source code is uploaded to GitHub.\footnote{\url{https://github.com/dlbac/DlbacAlpha/tree/main/understanding_dlbac_alpha}}

\subsection{Integrated Gradients}
\label{integratedGradients}
\textit{Integrated Gradients} is an effective interpretation technique that focuses on attributing the decisions of a neural network to the input features of the prediction samples.
It attributes a network's decision to its input features in terms of gradient, which specifies the most effective elements for a decision. To understand the \textit{decision} of \(\DLBACALPHA\) for a \textit{tuple}, it is required to provide the \textit{user/resource metadata values}, the \textit{decision}, and the \textit{neural network} as input to the \textit{Integrated Gradients}. Then, \textit{Integrated Gradients} outputs the \textit{attribution scores} of the input metadata that we normalize in the scale of 0 to 1, denoting the degree of impact on the decision. Such an interpretation is known as \textit{local interpretation} which helps to better understand each decision of a neural network.
It is also helpful to have a \textit{global interpretation}~\cite{gu2019semantics} of a network to understand the network's overall knowledge. Generally, it takes a set of decisions to generate a global interpretation of a network.

\begin{figure}[t]
    \vspace{-2ex}
    \centering
	\includegraphics[width= 1.0\linewidth, scale = 1.0]
	{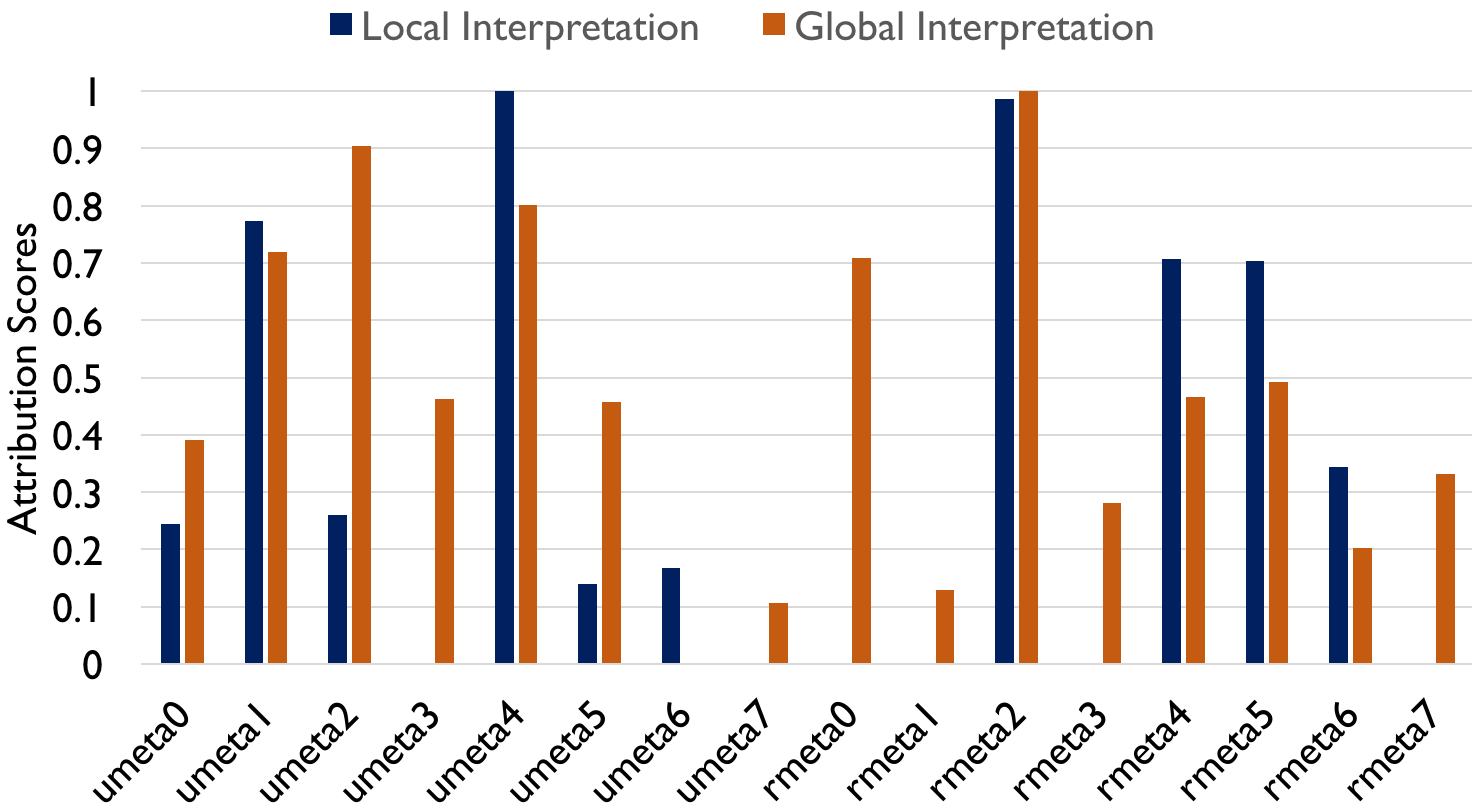}
	\caption{Local and Global Interpretation (blue: local interpretation of a decision in \(\DLBACALPHA\). orange: global interpretation of \(\DLBACALPHA\) for the grant access of an operation).}
	\label{fig:localGlobalInterpretation}
	\vspace{-2ex}
\end{figure}

For \(\DLBACALPHA\), we investigate explainability for both specific access control decision and the overall knowledge learned by the network. For this purpose, we train \(\DLBACALPHA\) on our \(\SMALLA\) dataset. Then, we request \textit{op1} operation access to \textit{projectD} resource for a user \textit{Dave} and the Decision Engine grants the request. To learn the reason behind this decision, we perform local interpretation with metadata values of \textit{Dave} and \textit{projectD}, the decision (\textit{grant} access on op1 operation), and the \textit{\(\DLBACALPHA\) network}. As depicted in Figure~\ref{fig:localGlobalInterpretation} (blue bars), for this particular request, user's \textit{umeta4} and resource's \textit{rmeta2} metadata are the most important and influential.

To achieve global interpretation for the \textit{grant} access to \textit{op1} operation, we take the \textit{op1} access information for a set of fifty random samples with grant access from the \(\SMALLA\) dataset. (Note that the more samples we use, the more precise the result we get. However, Integrated Gradients is memory inefficient, so we could not test more samples using our workstation with 16 GB of memory). We provide each user/resource metadata and their respective decisions for the \textit{op1} operation to \textit{Integrated Gradients} to obtain the global interpretation. Figure~\ref{fig:localGlobalInterpretation} (orange bars) depicts the global interpretation of \(\DLBACALPHA\) for the \(\SMALLA\) dataset for the \textit{grant} access to \textit{op1} operation. 
The result identifies resource's \textit{rmeta2} as the most influential metadata, user's \emph{umeta2} as second most influential metadata, and the \textit{attribution scores} of other metadata.

\subsubsection{Application of Integrated Gradients based Understanding}
\label{integratedGradientsJustification}
\begin{figure}[t]
    \vspace{-2ex}
    \centering
	\includegraphics[width=\linewidth]
	{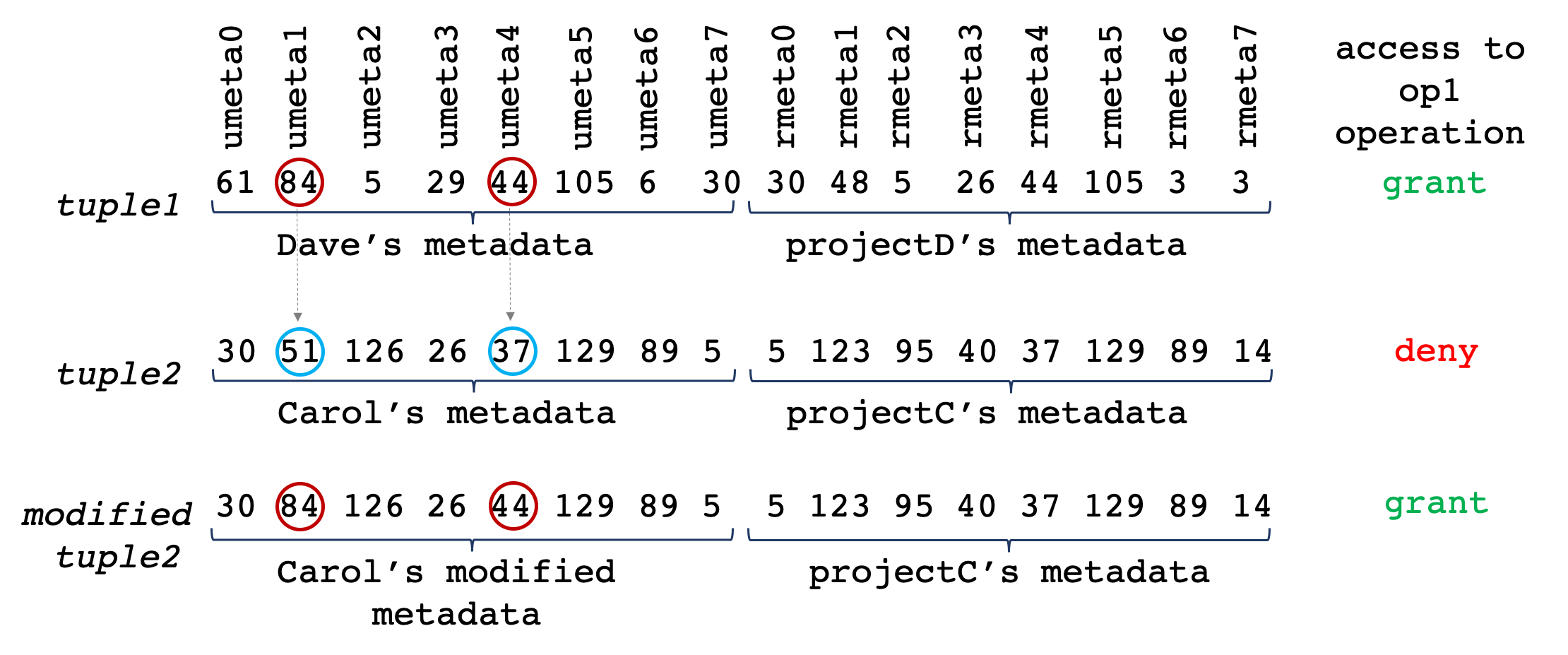}
	\caption{Applying Integrated Gradients to Grant Access.}
	\label{fig:modifyAccessThroughIG}
	\vspace{-2ex}
\end{figure}

Improved explainability could be utilized to achieve other benefits. 
For instance, developers can utilize interpretation techniques to debug~\cite{nguyen2015deep} incorrect decisions from the \textit{neural network} in \textit{decision engine}. We show that \textit{Integrated Gradients} based interpretation can be used to grant/deny access permissions by modifying proper metadata. As shown in Figure~\ref{fig:modifyAccessThroughIG} (tuple2), the user \textit{Carol} doesn't have op1 access on \textit{projectC} resource. Applying local interpretation on other tuple with op1 access (e.g., Dave has the op1 access to projectD resource as shown in Figure~\ref{fig:modifyAccessThroughIG} (tuple1)) revealed the \textit{attribution scores} of different metadata for op1 operation. As circled in tuple1, \textit{Dave}'s `umeta1' and `umeta4' are the most dominant metadata for this specific access. To grant Carol's op1 access on projectC resource, we utilize \textit{attribution scores} of tuple1. Our result shows that replacing Carol's `umeta1' and `umeta4' metadata value with Dave's metadata value enables Carol's op1 access to projectC resource. As reported, modifying one or two significant metadata changes corresponding access. The opposite holds for least-significant metadata where modifying multiple metadata could not alter related access. 

We also experiment with such a metadata value modification across all the samples of a specific decision (e.g., tuples with `deny' access on op1) in the same dataset to see the impact of a global interpretation.
The idea is to alter the metadata values of influential metadata for a specific decision. For instance, according to Figure~\ref{fig:localGlobalInterpretation}, the \emph{rmeta2} has the most influence on denying access. If we alter the \textit{rmeta2} value of any tuple, say \emph{tupleA}, with a rmeta2 value from a known tuple with \emph{grant} access, say \emph{tupleB}, the chance of denying access for \emph{tupleA} might reduce and increases the chance of getting access. Also, altering the value of multiple influential metadata of the same tuple may eventually help to get access.

\begin{figure}[t]
    \vspace{-2ex}
    \centering
	\includegraphics[width= 0.8\linewidth, scale = 0.8]
	{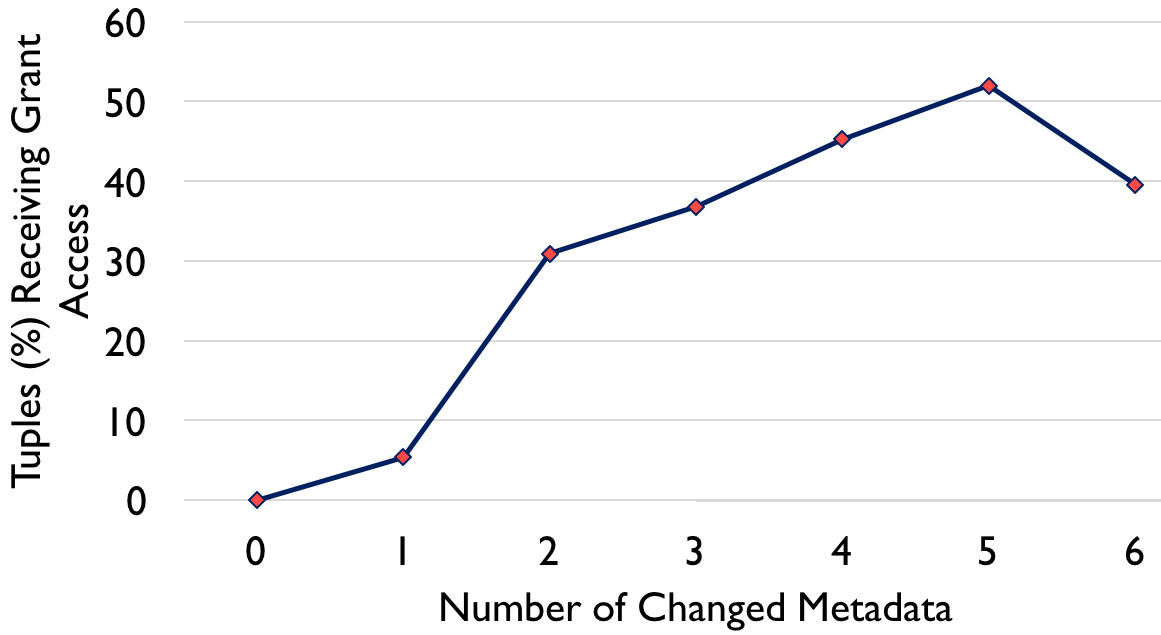}
	\caption{Modifying Metadata Value Based on Integrated Gradients Across Tuples (4581 tuples) with Deny Access.}
	\label{fig:modifyMetadataAcrossTuplesWithDeny}
	\vspace{-2ex}
\end{figure}

To investigate that we alter different metadata values, one by one, of all the tuples with \emph{deny} access (4581 such tuples) on \emph{op1} in order of their significance level in the global interpretation. For example, we first change the value of rmeta2 metadata, next umeta2 metadata, and so on. We utilize the user/resource metadata values from tuple1 in Figure~\ref{fig:modifyAccessThroughIG} as a known tuple with the `grant' access for \emph{op1} operation. As described in Figure~\ref{fig:modifyMetadataAcrossTuplesWithDeny}, initially, with no change, no tuple has been granted access. However, with the change of first metadata (rmeta2), around 5\% of the tuples receive grant access. By changing the second metadata value, around one-third of the tuples get grant access. A similar surge continues to obtain grant access for over half of all the tuples before decreasing for sixth metadata (rmeta5) modification, reaching below 40\%. It indicates that some tuples with such a big number of metadata modifications fall under a different distribution for which they have the `deny' access. Overall, using this technique, a system admin can estimate the impact of metadata value vs. dependent accesses.

\begin{figure}[t]
    \vspace{-1ex}
    \centering
	\includegraphics[width=\linewidth]
	{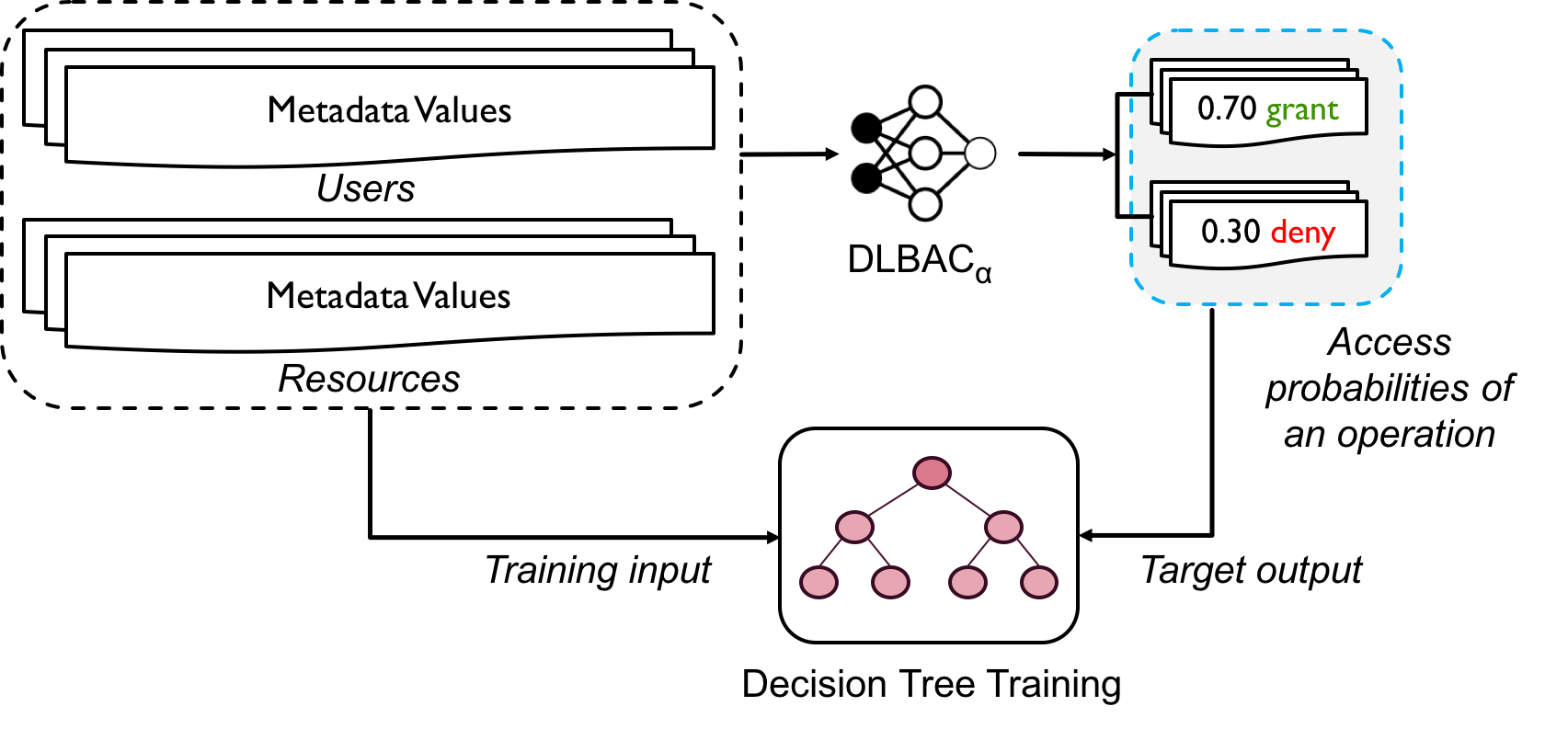}
	\caption{Knowledge Transferring Technique.}
	\label{fig:ktTechniqueDlbac}
	\vspace{-2ex}
\end{figure}

\subsection{Knowledge Transferring}
\label{knowledgeTransferring}
Although \textit{Integrated Gradients} determines the \textit{attribution scores} of each metadata for a decision, it does not establish the relationship among metadata or the network's logic~\cite{sundararajan2017axiomatic}. \textit{Knowledge Transferring} could help to identify such relationships. With \textit{Knowledge Transferring}, we can extract a decision tree to approximately understand the decision of \(\DLBACALPHA\) in the form of traditional rules. While accurate reconstruction of the representation details is infeasible, the generated decision tree will give an approximate explanation of the underlying logic of \(\DLBACALPHA\).
Though the decision tree makes classification decisions understandable~\cite{scikit-learn}, it does not generalize as well as deep neural networks. Interestingly, a neural network's generalization ability is likely to be transferred to a decision tree through a method called \textit{distillation}~\cite{hinton2015distilling}, which has been widely used in ML literature and we adopt in this paper.
As explained in Section~\ref{sec:organizingTrainingData}, \(\DLBACALPHA\)
outputs whether a user has access to any resource by giving the \textit{probability} of granting the access instead of a direct \textit{yes/no} answer. Therefore, we can determine those \textit{probability outputs} for all the \textit{tuples} in a dataset and train a decision tree, as shown in Figure~\ref{fig:ktTechniqueDlbac}. These probabilities represent the \textit{knowledge} of the neural network, and we aim to transfer it to a decision tree. 

To explore \textit{Knowledge Transferring} technique, we train \(\DLBACALPHA\) for the \(\SMALLA\) dataset that we used for the Integrated Gradients experiment. We take \textit{op1} access \textit{probabilities} for all the samples. Then, we train a decision tree (DT) on the same dataset. However, instead of giving corresponding ground-truth permissions from the dataset, we provide the \textit{probability} outputs of \(\DLBACALPHA\). Eventually, we construct a DT 
with a maximum depth of eight that facilitates retrieving underlying rules for any specific decision. While the tree serves the global interpretation, a rule for any specific decision represents the local interpretation. 
We apply the metadata values of \textit{Dave} and \textit{projectD} that we described in previous section, and retrieve the access rules from the DT for \textit{op1} operation. We observe that \textit{Dave} obtained \emph{grant} access to \textit{projectD} for \textit{op1} based on following rule:
$\bigg\langle$
$\Big\langle$
$\langle$
umeta0 $>$ 31 $\land$ umeta0 $<$ 63
$\rangle$
$\land$
$\langle$
umeta2 $<$ 20
$\rangle$
$\land$
$\langle$
umeta6 $<$ 50
$\rangle$
$\Big\rangle$
$\land$
$\Big\langle$
$\langle$
rmeta0 $<$ 72
$\rangle$
$\land$
$\langle$
rmeta2 $<$ 18
$\rangle$
$\land$
$\langle$
rmeta5 $<$ 111
$\rangle$
$\Big\rangle$
$\bigg\rangle$.
It is worth mentioning that the decision tree should be generated with unlimited depth to obtain more precise rules.
Note that the Integrated Gradients and Knowledge Transferring techniques are orthogonal in terms of insights they each provide into the neural network and do not substitute each other. For better insights, one could use both methods.

%% file: contents/discussions.tex
\section{Future Research Directions} \label{sec:discussion}

In this section, we discuss some of the challenges for DLBAC and explore some ideas of how those could be addressed.

\textbf{Access Control Administration.} A major task in access control is policy administration, i.e. updating the system's rules to affect a particular change. It is a significant challenge for DLBAC as a policy change amounts to adjusting the current neural network's weights. This can, of course, be obtained by retraining the network. However, that is neither ideal nor cost-effective. Fine-tuning is one of the common approaches that helps a neural network to learn new changes by updating the current network's weight~\cite{kaya2019analysis}. This allows one to implement policy changes by making minor changes to the network. However, the network should handle the issue of catastrophic forgetting, which is a common pitfall in fine-tuning. 
We believe these issues can be effectively mitigated by developing methods based on life-long learning~\cite{parisi2019continual} for DLBAC.

\textbf{Adversarial Attacks.} Adversarial attacks are a common concern for any machine learning based system, deep learning networks in particular. An adversary can obtain unauthorized access by fooling the network with modified samples that are indistinguishable from natural ones by human~\cite{zheng2019distributionally}. 
However, such attacks could be mitigated by applying adversarial training~\cite{madry2017towards}.
In the context of DLBAC, an adversarial attack is to acquire access permissions based on modified/perturbed user and resource metadata. 
In access control, the datasets and the adversarial attack profile are somewhat interesting and different from traditional image domains. One typically expects a mix of categorical and continuous metadata. Moreover, since some metadata are more trustworthy than others, an adversary does not have the complete flexibility to change an entire sample imperceptibly.
These observations could be leveraged to develop better defenses against adversarial attacks in DLBAC. Another related aspect needs to investigate whether DLBAC can efficiently handle an access request if some of its user/resource metadata are deleted, a.k.a attribute hiding attack.

\textbf{Bias and Fairness.} As DLBAC learns based on metadata distributed across various parts of an enterprise, there might be different types of human biases or errors in training data.
As such, the DLBAC network trained based on such data can inadvertently be biased, favoring some decisions. For example, as observed in the \(\AMAZONKAGGLE\) dataset, most of the authorization tuples were with `grant' decision, and \(\DLBACALPHA\) instances were biased towards the same decision. Certain metadata could be influenced by various factors including ethnographic.
Therefore, to obtain a fair and trustworthy DLBAC system, it is critical to audit training data, evaluate decisions for fairness, and establish a proper feedback loop~\cite{mehrabi2019survey}.

\textbf{DLBAC in Tandem with Traditional Access Control.} Evidently, in practice, we do not foresee (nor advocate) that DLBAC will simply substitute traditional forms of access control immediately. One research challenge is how DLBAC could be effectively integrated to operate in tandem with traditional access controls such as RBAC or ABAC. One of the issues that will arise is conflicting decisions between, say, RBAC and DLBAC. If that conflict were to be resolved in favor of RBAC, that decision could be used to fine-tune the DLBAC network. DLBAC could also be used in the background for monitoring or auditing purposes.




%% file: contents/conclusion.tex
\section{Conclusion} \label{sec:conclusion}
We proposed DLBAC, a deep learning based access control approach, to deal with issues in classical access control approaches. As DLBAC learns based on metadata, it obviates the need for attribute/role engineering, policy engineering, etc. 
We implemented \(\DLBACALPHA\), a prototype of DLBAC, using both real-world and synthetic datasets.
We demonstrated \(\DLBACALPHA\)'s effectiveness as well as its generalizability. 
As the core of DLBAC is a neural network, we applied two different state-of-the-art techniques to understand DLBAC decisions in human terms.
We also discussed some future directions to build new models upon DLBAC and address current challenges, including access control administration issues.
